\documentclass[runningheads]{llncs}

\usepackage{eccv}

\usepackage{eccvabbrv}

\usepackage{graphicx}
\usepackage{booktabs}

\usepackage[accsupp]{axessibility}
\usepackage[textsize=tiny]{todonotes}
\usepackage{soul}
\usepackage{cuted}
\usepackage{diagbox}
\usepackage{amsmath}
\usepackage{amssymb}
\usepackage{mathtools}
\usepackage{multirow}
\usepackage{enumitem}
\usepackage{microtype}
\usepackage{graphicx}
\usepackage{subcaption}
\usepackage{booktabs}
\usepackage{subcaption}

\newcommand*{\samethanks}[1][\value{footnote}]{\footnotemark[#1]}

\usepackage{hyperref}

\usepackage{orcidlink}

\begin{document}

\title{Enhanced Neural Video Representation Compression Across Extreme Complexity and Quality Scales}

\titlerunning{Enhanced Neural Video Representation Compression}

\author{Ho Man Kwan\inst{1}\thanks{Equal contribution.}\orcidlink{0000-0002-8283-4513} \and
Tianhao Peng\inst{1}\samethanks\orcidlink{0009-0008-5294-2880
} \and
Fan Zhang\inst{1}\orcidlink{0000-0001-6623-9936} \and
Mike Nilsson\inst{2}\orcidlink{0000-0002-9134-6546} \and \\
Andrew Gower\inst{2}\orcidlink{0009-0007-6136-0949} \and
David Bull\inst{1}\orcidlink{0000-0001-7634-190X}}

\authorrunning{H. M.~Kwan et al.}

\institute{Visual Information Lab, University of Bristol, UK \\
\email{\{hm.kwan, tianhao.peng, fan.zhang, dave.bull\}@bristol.ac.uk} \and
Network Connectivity Services Research, BT, UK \\
\email{\{mike.nilsson, andrew.p.gower\}@bt.com}
}

\renewcommand{\thefootnote}{\fnsymbol{footnote}}
\maketitle
\renewcommand{\thefootnote}{\arabic{footnote}}
\setcounter{footnote}{0}

\begin{abstract}
Implicit neural representations (INRs) have recently emerged as a promising approach to video compression, delivering competitive rate-distortion performance alongside rapid decoding. However, existing neural video codecs struggle to balance complexity and scalability. Lightweight models often suffer from degraded compression performance when scaled to different bitrate/quality levels, whereas high-performance models exhibit limited scalability, as their model complexity typically increases with quality. This lack of a unified architecture capable of maintaining consistent complexity across a wide range of bitrates severely limits their diverse real-world deployment.
To address these challenges, we introduce NVRC++\footnote{Project page: \url{https://hmkx.github.io/nvrc-pp/}}, a novel INR-based video codec that utilizes a lightweight INR with multiple high-resolution feature grids, providing high scalability at any given complexity level. This is paired with an optimization framework that enables efficient overfitting on high-resolution grids for long video sequences, thereby exploiting spatio-temporal redundancies without prohibitive computational or memory overhead. Additionally, an advanced entropy model is designed for efficiently compressing the high-dimensional grid parameters. As a result, NVRC++ provides four complexity levels (from 7kMACs/pixel to 360kMACs/pixel),  each spanning wide bitrate and quality ranges while supporting real-time decoding. The experimental results show that NVRC++ offers a much faster decoding speed (up to 8.5x) compared to the SOTA INR-based video codec, NVRC, while delivering comparable performance.
\end{abstract}

\section{Introduction}
\label{sec:intro}

Recent work in neural video compression has delivered coding performances that are comparable to the latest standard-based video codecs \cite{bull2021intelligent}. Many of these learning-based methods \cite{lu2019dvc,li2024neural} are based on autoencoder networks, which transform input videos into a latent domain before performing quantization and entropy coding. Such approaches typically learn a generalized model from large training datasets and are deployed during the inference phase to compress an unseen video. Due to the large capacity required for a generalized model, these methods are often associated with high computational complexity, which hinders their application in real-world scenarios.

More recently, implicit neural representations (INR) \cite{chen2021nerv, gomes2023video, kwan2024hinerv, kwan2024nvrc} have provided an attractive alternative solution for video compression. Typically, an INR-based video codec utilizes an overfitted neural network to represent a video and further employs model compression and/or entropy coding techniques to encode model parameters, including layer weights \cite{chen2021nerv, li2022nerv} and feature grids \cite{chen2023hnerv, lee2023ffnerv, kwan2024hinerv, kim2023c3}. Although it is relatively slow when involving an overfitting process in encoding, these INR-based methods can effectively exploit spatio-temporal redundancies in an implicit manner via the shared neural network layers that are subsequently used for decoding video frames. The latest neural representation based codecs \cite{kwan2024nvrc} have shown competitive rate-quality performance compared to conventional codecs such as HEVC HM \cite{hm} and VVC VTM \cite{vtm}, as well as the best performing autoencoder-based neural video codecs \cite{li2022hybrid, li2023neural}.

While demonstrating promising compression performance, a critical gap remains in the research community with lightweight and high-performance INR-based/overfitted codecs. Specifically, most performance focused methods \cite{chen2021nerv, li2022nerv, chen2023hnerv, kwan2024hinerv, kwan2024nvrc} suffer from poor scalability, as they require different model configurations for various target quality levels or rate points. Consequently, their computational complexity increases with video quality, resulting in significant decoding computational overhead at high bitrates. This low scalability characteristic prevents their application in real-world scenarios, while a practical neural video codec should ideally achieve a wide bitrate/quality range with a single-size model. Conversely, lightweight overfitted codecs can achieve rate scalability with a single or narrow complexity range, but they yield inferior compression performance due to their lightweight models \cite{leguay2024cool, kim2023c3, kwan2025ultra} and the independent coding of patches \cite{kim2023c3}. There are also overfitted INR-based codecs which do not benefit much from the overfitting process \cite{gao2025pnvc}, which can be effective for exploiting global spatial-temporal redundancy \cite{kwan2024hinerv, kwan2024nvrc}. This lack of a unified architecture capable of reaching diverse complexity levels while maintaining high rate-quality scalability prevents their application in diverse real-world scenarios.

An efficient way to bridge this complexity and scalability gap is the adoption of high-dimensional feature grids \cite{leguay2024cool, kim2023c3, kwan2025ultra} for the high performance codec, but with diverse complexity settings. By storing dynamic spatio-temporal features directly, these grids allow an INR to represent higher-quality details without increasing model complexity. However, directly deploying high-dimensional feature grids can be suboptimal. While recent work shows that high performance INR-based codecs offer optimal performance when encoding an entire long video sequence with a single model \cite{kwan2024hinerv, zhang2024boosting, zhu2025msnerv}, optimizing INRs with high-resolution grids for long sequences requires substantial computational and memory resources, particularly when the interdependency between grid features must be modeled, for instance, by an entropy model. To manage this, existing works typically use high-resolution grids but encode the video in individual patches \cite{kim2023c3} or short video clips \cite{kwan2025ultra, gao2025pnvc}. Furthermore, while high-resolution grids improve performance at high bitrates, they may impact low-bitrate performance because these grids increase the data volume, resulting in sub-optimal compression performance \cite{kwan2025ultra}.

To address the scalability issue and bridge the performance gap across the entire complexity spectrum, this paper proposes a novel neural video representation compression framework, NVRC++. Targeting high scalability, it utilizes a neural representation consisting of multiple high-resolution feature grids, allowing a single model to achieve a wide bitrate/quality range and fast decoding with every complexity configuration.
To avoid suboptimal performance when directly deploying high-resolution grids (as mentioned above), we also developed an enhanced optimization framework inspired by the hierarchical coding structure in conventional video codecs to allow the optimization of these high-dimensional grids under computational resource constraints. This improves performance across the entire bitrate/quality range and the decoding speed. Furthermore, to reduce the additional storage required by high-resolution grids, NVRC++ is integrated with an improved multi-dimensional grid entropy model to better exploit spatio-temporal redundancy within videos and between grids of different scales. Finally, our framework is built upon HiNeRV++, an improved INR designed for better performance and efficiency by leveraging multi-reference context from higher-resolution feature grids. The main contributions are summarized:

\begin{figure}[t!]
    \centering
    \scriptsize

    \includegraphics[width=\linewidth]{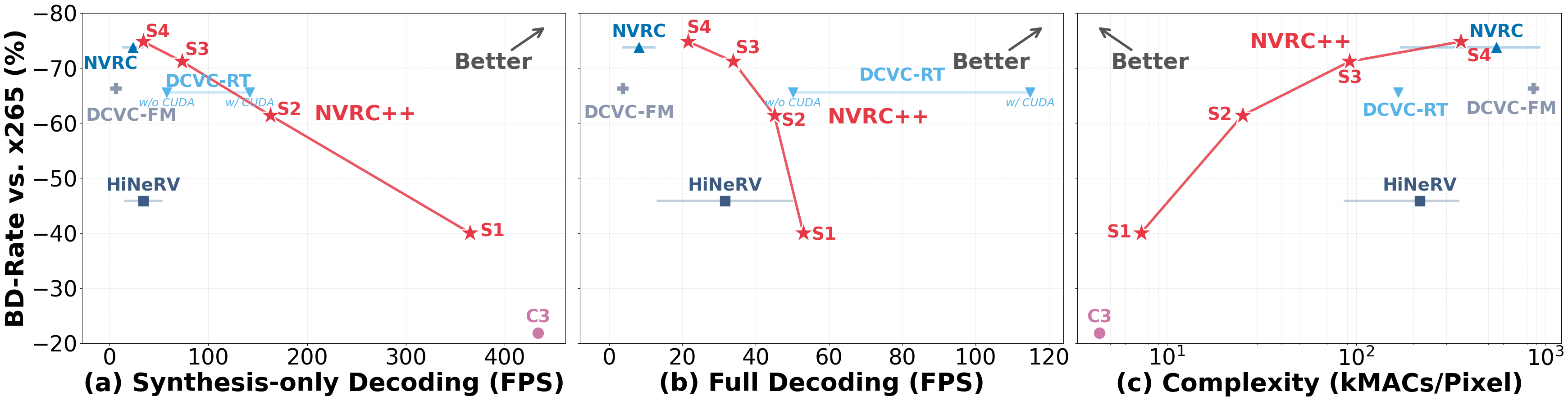}
    \caption{\footnotesize The trade off between decoding complexity (in terms of FPS and MACs/pixel) and coding performance (in BD-rate against x265 (\textit{veryslow})) \cite{x265} of the proposed method and existing neural video codecs. \textbf{(a)} BD-rate vs. \textbf{\emph{synthesis-only}} decoding speed, where for over-fitted approaches the entropy decoding step is excluded as it is a process separate from frame decoding \cite{kim2023c3, kwan2024nvrc}; \textbf{(b)} BD-rate vs. \textbf{\emph{full}} decoding speed, where the entropy decoding step is included (C3 \cite{kim2023c3} is omitted here as no entropy encoder/decoder is available for it, and for DCVC-RT \cite{jia2025towards} we report the speed both with and without its custom CUDA kernels); \textbf{(c)} BD-rate vs. computational complexity in kMACs/pixel. NVRC++ achieves a superior complexity-compression trade-off, delivering state-of-the-art performance across an unprecedented quality range at every complexity level from 7 to 360 kMAC/pixel. Note that we have not utilized custom CUDA kernel, while still achieving very fast decoding speed.}
    \label{fig:tradeoff}
    \vspace{-15pt}
\end{figure}

\begin{itemize}[leftmargin=*]
    \item \textbf{A novel INR-based codec with high scalability} that supports multiple complexity levels. By utilizing high-resolution features, it achieves high scalability and fast decoding at each complexity level while maintaining competitive coding performance.

    \item \textbf{An efficient optimization framework} inspired by the hierarchical coding structures in conventional video codecs; it effectively exploits spatio-temporal redundancies within parameters, enables the optimization of high dimensional grids under computational constraints, and achieves lower decoding latency. 

    \item \textbf{An advanced multi-dimensional grid entropy model}, which offers an efficient compression method for high dimensional grid parameters by leveraging multiple prior information, yielding high entropy encoding and decoding speeds.

    \item \textbf{An enhanced INR, HiNeRV++}, which supports real-time decoding across various complexity configurations, achieving a wide bitrate range within each configuration and contributing to superior compression performance.
\end{itemize}

The NVRC++ codecs (four variants at different complexity levels) have been benchmarked against state-of-the-art conventional, autoencoder-based, and INR-based video codecs. The results (as summarized in \cref{fig:tradeoff}) 
show that all NVRC++ codecs offer an excellent trade off between decoding complexity and coding performance. Specifically, NVRC++ (S3) achieves a much faster decoding speed (8.5 times) compared to the SOTA INR-based codec, NVRC, while maintaining comparable performance.

\section{Related Work}
\label{sec:related_work}

\noindent\textbf{Neural video compression}
\label{subsec:nvc}
has emerged as a prominent research field in recent years, leveraging deep learning techniques to achieve improved coding efficiency. Based on the variational autoencoder framework and end-to-end optimization that originate from learned image compression \cite{balle2018variational}, DVC \cite{lu2019dvc} was the first end-to-end optimized video codec to adapt a residual coding-based architecture for traditional codecs, where each component is replaced by its learnable alternative. Later works achieved enhanced coding efficiency by integrating a variety of techniques such as 3D modeling \cite{habibian2019video}, conditional coding \cite{liu2020conditional,ladune2021conditional, ho2022canf,li2024neural}, conditional residual coding \cite{chen2024maskcrt} and Transformer-based architectures \cite{mentzer2022vct, xiang2022mimt,chen2023neural}. The SOTA neural video codecs have demonstrated superior capability \cite{teng2024benchmarking,jia2025towards,jiang2025ecvc,jiang2025biecvc,sheng2025bi
}, offering performance comparable to the latest standard video codecs \cite{sullivan2012overview, bross2021overview}.

\vspace{5pt}\noindent\textbf{Neural representation for video compression.}
When applied to video coding, an INR \cite{sitzmann2020implicit, mescheder2019occupancy, park2019deepsdf} learns a mapping between coordinate inputs and the corresponding target pixel values for a single instance of video \cite{sitzmann2020implicit, li2022nerv} or a set of videos \cite{he2023towards}. This approach effectively converts video compression into a model compression task, where the network represents videos with model weights. The pioneering work NeRV \cite{chen2021nerv} established a three-step compression pipeline: model pruning with fine-tuning to reduce model size, weight quantization to lower precision, and entropy coding to minimize statistical correlations. Subsequent work focused on enhancing different stages/components in this pipeline, including entropy minimization \cite{gomes2023video,maiya2023nirvana,kwan2024immersive,zhang2024boosting}, architectural innovations \cite{li2022nerv, zhu2025msnerv}, and flow-based motion compensation \cite{zhang2021implicit,lee2023ffnerv,he2023towards}. To further improve coding efficiency, single/multiple resolution feature grids were introduced to effectively represent adaptive visual content information in the latent space \cite{lee2023ffnerv,kim2023c3,leguay2024cool}. A significant milestone was the NVRC framework \cite{kwan2024nvrc}, which proposed an end-to-end optimized compression framework for neural video representations. This has been reported to be the best INR-based video codec to date, offering comparable performance to VVC VTM (Random Access mode).

\section{Method}
\label{sec:method}

\noindent\textbf{Preliminaries.}
In the context of INR-based video compression, most methods utilize an overfitting approach initialized by NeRV \cite{chen2021nerv}; i.e., neural representations are used for coding videos in an instance-adaptive manner, where a neural network is trained from scratch to learn the coordinate-to-output mapping. Specifically, given a patch coordinate $(i, j, k)$, where $0 \leq i < \frac{W}{W_{patch}}$, $0 \leq j< \frac{H}{H_{patch}}$, and $0 \leq k < \frac{T}{T_{patch}}$, of a 3D patch $\hat{V}_{patch}$ with size $T_{patch} \times H_{patch} \times W_{patch}$ within the input video $V^{gt}\in \mathbb{R}^{T\times H\times W \times C}$, the corresponding normalized pixel coordinate $x$ within that patch is first embedded into high dimensions, which is commonly achieved through positional encodings \cite{mildenhall2021nerf, chen2021nerv}:
\begin{equation}
    \gamma(x) = \text{Concat}\big(\sin(2^0 \pi x), \cos(2^0 \pi x), \dots, \sin(2^{L_{freq}-1} \pi x), \cos(2^{L_{freq}-1} \pi x)\big),
\end{equation}
where $L_{freq}$ is the number of frequency bands. Alternatively, recent works utilize feature grids \cite{lee2023ffnerv, ladune2023cool, kim2023c3, leguay2024cool, kwan2024hinerv}, where a set of learnable feature embeddings $\mathcal{F}$ is stored in multi-resolution grids. The corresponding feature $f$ is extracted via trilinear interpolation at each level and concatenated across levels:
\begin{equation}
    f = \text{Concat}\big(\text{Interp}(x, \mathcal{F}_0), \dots, \text{Interp}(x, \mathcal{F}_{L_{grid}-1})\big),
\end{equation}
where $\mathcal{F}_l$ represents the grid at resolution level $l$, and $L_{grid}$ is the number of grid levels. Typically, these grids are implemented with multiple resolutions, where the grid dimensions are significantly lower than the original video dimensions to maintain a compact representation. By performing a forward pass at every coordinate using the embedded features, a reconstruction, $V$, of the video $V^{gt}$ can be decoded. The neural network parameters $\theta$ can thus be treated as a compact representation of the signal, and techniques such as quantization and entropy coding \cite{chen2021nerv, gomes2023video} are then used to further compress $\theta$ into a bitstream. This process is typically lossy, and the quality is subject to both the representational capability and the compression ratio. The objective of INR-based video coding can thus be considered as a classical rate-distortion optimization problem based on neural compression \cite{balle2017end}:
\begin{equation}
    \mathrm{argmin}_\theta \mathcal{L} = \mathrm{argmin}_\theta \left(R(\hat{\theta}) + \lambda D(V, V^{gt})\right).
\end{equation}
Here, $\theta$ and $\hat{\theta}$ are the original and quantized model parameters. $R$ and $D$ are the measurements of rate and distortion, respectively.

While the latest INR-based video codecs demonstrate promising compression performance \cite{chen2023hnerv, kwan2024nvrc, zhang2024boosting}, there is a major limitation that hinders practical applications: their model sizes typically vary with bitrate and quality. This implies that the model complexity scales with the target bitrate/quality, which is a critical problem: a large INR-based codec \cite{kwan2024hinerv,kwan2024nvrc} could be as costly as the end-to-end autoencoder-based models \cite{lu2019dvc, li2021deep}, which are trained as a generic model. In this case, overfitted INR-based codecs could lose their natural advantage in low decoding complexity due to instance adaptive coding. Moreover, having a decoder with fixed or nearly fixed complexity could be beneficial for both software and hardware implementations.

Compared to NeRV-based methods, there are also approaches that utilize lightweight neural networks together with high-resolution feature grids, e.g., COOL-CHIC \cite{leguay2024cool} and C3 \cite{kim2023c3}, for encoding videos. These approaches can achieve a wide range of quality with a single representation due to the use of high-resolution features, which alleviates the need for a high-capacity decoder network to represent the fine-detailed visual features. However, these approaches are significantly outperformed by the best performing NeRV-based methods \cite{kwan2024hinerv, kwan2024nvrc} in terms of compression performance. This may be due to (i) the tiny decoders or entropy models used \cite{leguay2024cool, kim2023c3}, which limit expressivity for exploiting the spatio-temporal redundancies within signals and modeling the data distributions, and (ii) the strategy that partitions videos into independent chunks for coding \cite{kim2023c3}, which allows training with high dimensional feature grids within the limited hardware memory but overlooks the redundancy across chunks, as the chunks are independently coded.

\subsection{NVRC++}
\label{subsec:NVRC++}

\begin{figure*}[!t]
    \centering
    \includegraphics[width=1\linewidth]{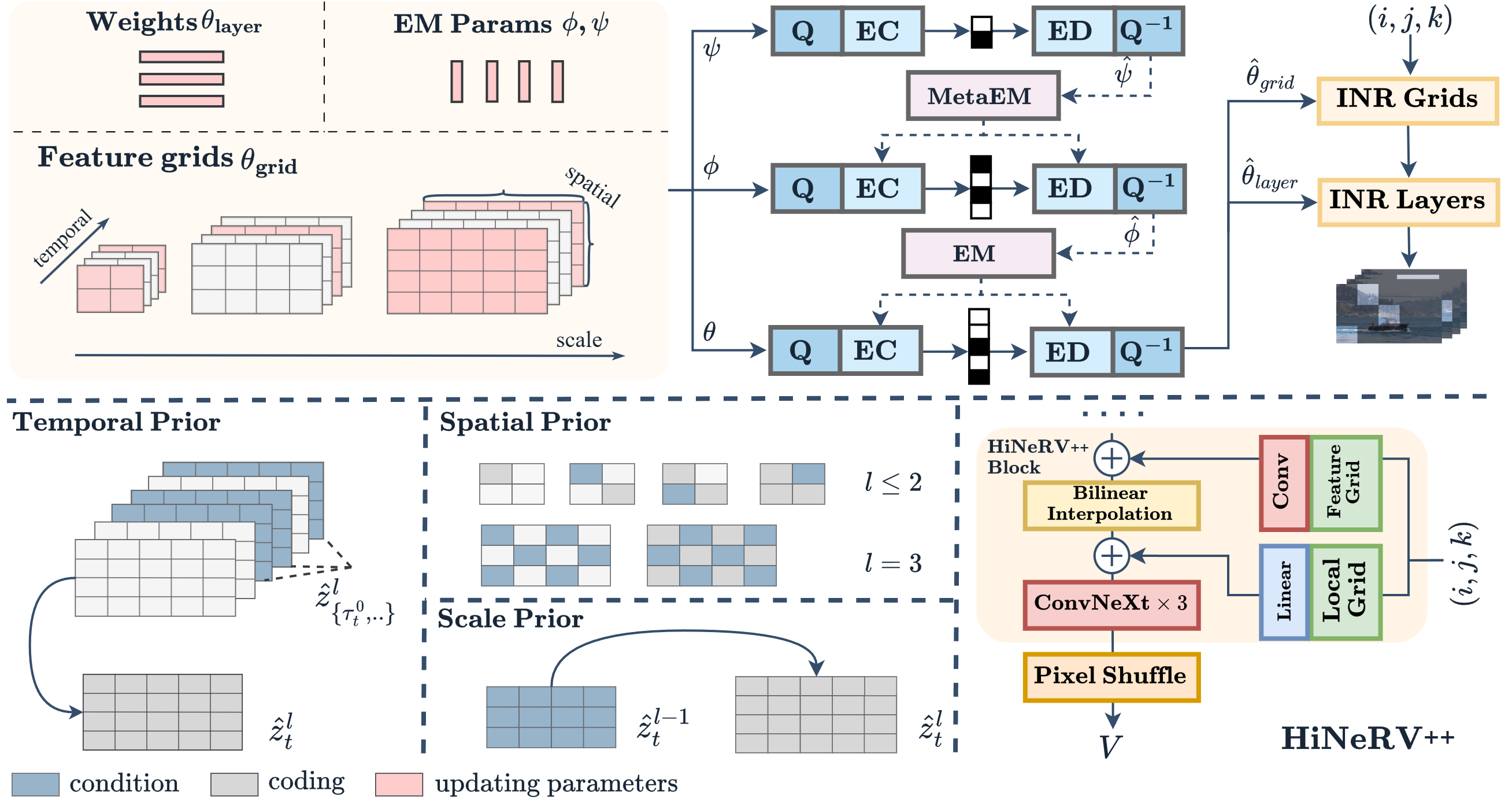}
    \caption{The overall NVRC++ framework. It contains a hierarchical parameter coding structure, while also contains an advance grid entropy model based on temporal, spatial and scale priors, and the improved INR network, HiNeRV++. It also features random sampling for updating only partial parameters for more efficient training process.}
    \label{fig:main}
    \vspace{-15pt}
\end{figure*}

In this context, we propose NVRC++, a novel neural video representation compression framework (shown in \cref{fig:main}) that advances INR-based video compression. Our framework significantly extends NVRC \cite{kwan2024nvrc}, a state-of-the-art INR-based codec, with four key innovations. 
(i) NVRC++ utilizes INRs with higher-resolution feature grids to decouple model complexity from reconstruction quality. This allows a single model configuration to span a wide bitrate range at fixed complexity, enabling adaptation to diverse hardware constraints; (ii) an enhanced optimization framework is employed to facilitate the training of the INR with high dimensional grids while performing compression at the same time; (iii) improved entropy models are designed to exploit the spatial and temporal redundancies between grid parameters more effectively; and (iv) an enhanced lightweight INR, HiNeRV++, which supports high reconstruction quality at different complexities while achieving fast decoding.

Following NVRC \cite{kwan2024nvrc}, in NVRC++, an INR with parameters $\theta = \{\theta_{grid}, \theta_{layer}\}$ is used for video coding, where $\theta_{grid}$ and $\theta_{layer}$ are the feature grids and network layer parameters, respectively. These parameters are optimized for reconstructing the video with the rate-distortion constraint, where the corresponding quantized parameters $\hat{\theta} = \{\hat{\theta}_{grid}, \hat{\theta}_{layer}\}$ are coded into the bitstream. Optimizing these parameters also involves quantization and entropy model parameters $\phi$ and their quantized parameters $\hat{\phi}$, as well as the meta compression parameters $\psi$ (and $\hat{\psi}$), where all these parameters form a hierarchical parameter structure and are jointly optimized in an end-to-end manner. We follow \cite{kwan2024nvrc} for the model compression framework, but with an improved optimization process (\cref{subsec:overfitting}), a grid entropy model with multiple priors (\cref{subsec:gridcoding}), and an enhanced lightweight INR (\cref{subsec:HiNeRV++}).

\subsection{Overfitting with high-resolution grid features}
\label{subsec:overfitting}

Although deploying INRs with high resolution grids offers better scalability, naively applying them will introduce higher complexity and a larger memory footprint, particularly when integrated with the contextual entropy models required for competitive compression. Specifically, when an autoregressive model is employed as the entropy model using techniques such as 3D convolutions \cite{kim2023c3}, all dependent parameters are processed simultaneously within a large grid tensor. This results in substantial computational and memory overhead. We address this limitation by introducing an \textit{in-parameter coding structure}. This structure models the dependency in a manner following conventional coding structures while making optimization based on overfitting feasible. In addition, we found that the model performance drops, especially at low bitrates, when the high resolution grids are introduced to the INR model directly. We propose to use a \textit{feature grid masking} technique, which regularizes and improves model performance.

\begin{figure*}[!t]
\centering
\begin{minipage}[t]{0.385\linewidth}
    \centering
    \includegraphics[width=\linewidth]{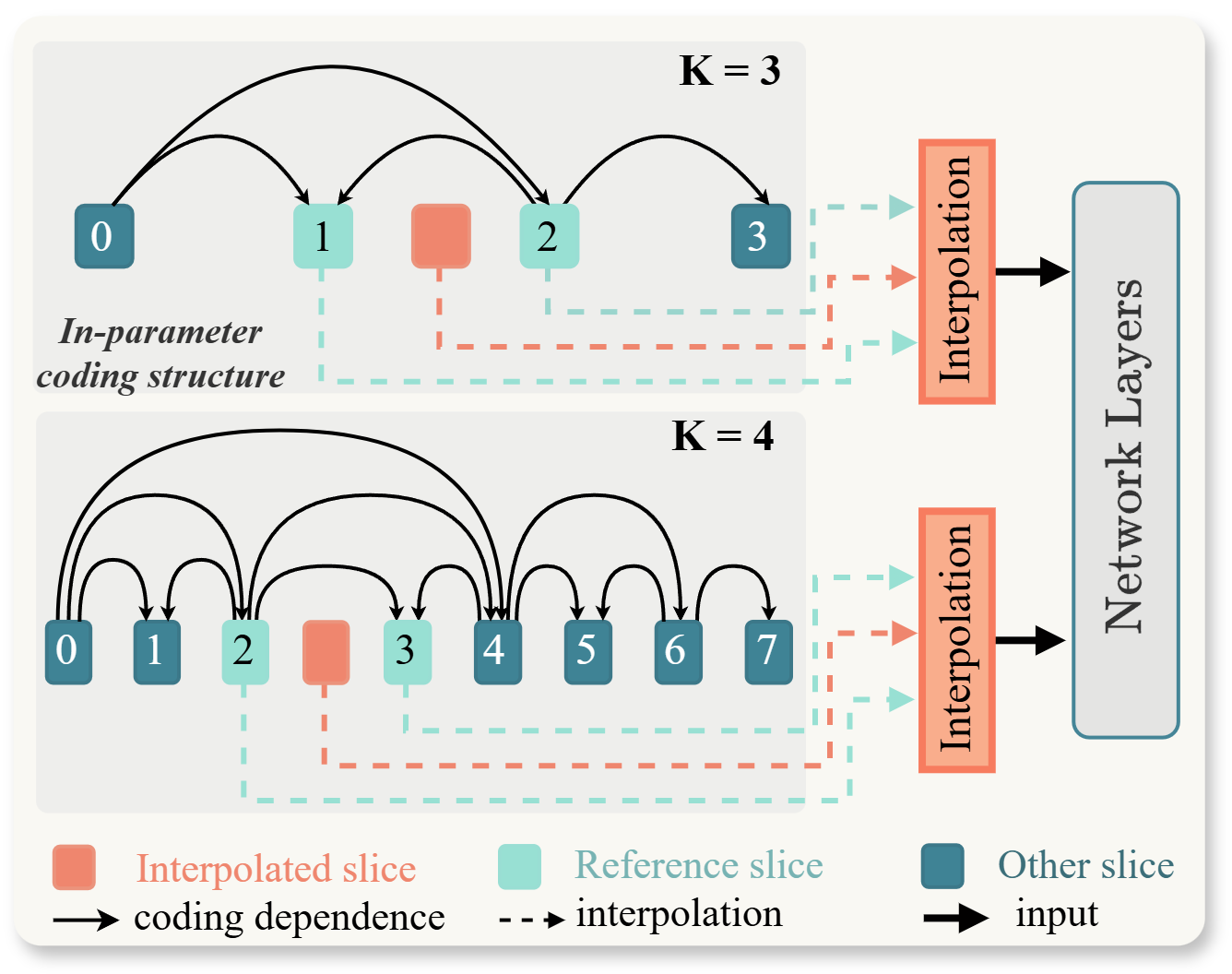}
\end{minipage}
\hfill
\begin{minipage}[t]{0.60\textwidth}
  \centering
\includegraphics[width=\linewidth]{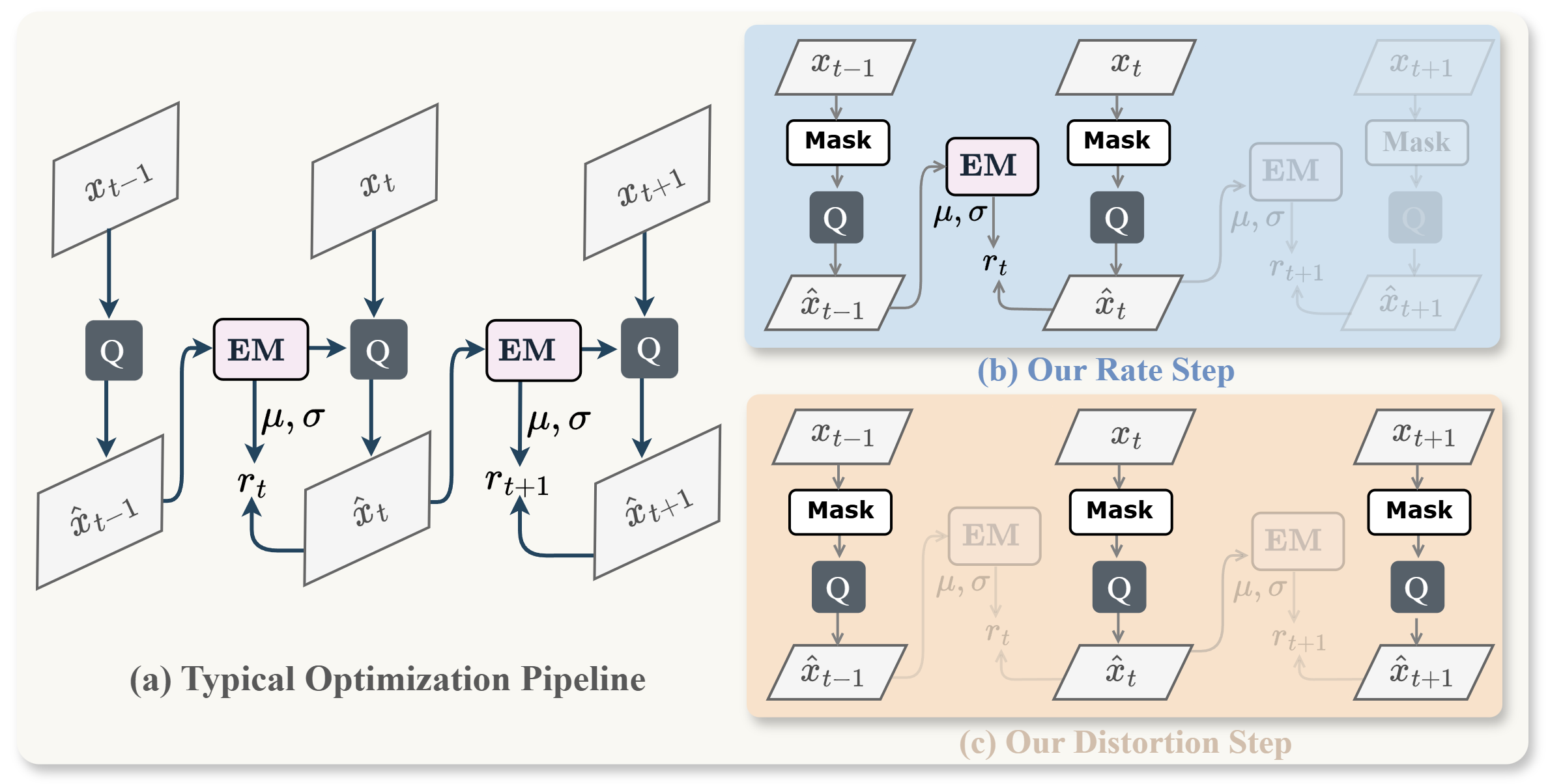}
\end{minipage}
 \caption{(\textbf{Left}) Proposed in-parameter coding and multi-reference inputs. Grid parameters are organized into temporal slices and compressed using a Hierarchical-B structure to leverage inter-slice dependencies. Unlike standard practices that rely solely on interpolated features, HiNeRV++ utilizes neighboring raw slices as multi-reference context to enhance reconstruction. While $K=3/4$ is shown for illustration, our implementation scales up to $K=10$. (\textbf{Right}) Optimization pipeline. To avoid the propagation issues of typical coding structure in overfitted codecs, we: (1) use non-conditional quantization to decouple slice optimization; (2) apply decoupled rate-distortion steps with random sampling (sampled components are highlighted); and (3) employ feature masking to improve the optimization with high-resolution grids. Low-delay coding is shown for demonstration, though Hierarchical-B is used in practice.}
 \label{fig:coding_structure}
  \vspace{-15pt}
\end{figure*}

\vspace{5pt}\noindent\textbf{In-parameter coding structure.} Inspired by the coding structures used in both conventional and neural video coding, we introduce an \emph{in-parameter coding structure} that partitions the feature grid parameters into temporal slices and performs coding by leveraging the dependency. 
Specifically, for a single resolution grid parameters $z \in \mathbb{R}^{T_{\text{grid}} \times H_{\text{grid}} \times W_{\text{grid}} \times C_{\text{grid}}}$ (and the corresponding quantized parameters \(\hat{z}\)), we divide it into temporal slices $\{z_t \mid 0 \le t < T_{\text{grid}}\}$ and utilize a contextual entropy model for modeling the conditional distribution $p(\hat{z}_{t}|\hat{z}_{\tau^{0}_{t}}, ...)$, which is subject to the rate–distortion objective. Here, we utilize the hierarchical B-frame coding structure with $K$ levels (an example is shown in \cref{fig:coding_structure} (Left)), where the number of temporal slices at each level is $1, 1, \dots, 2^{K-2}$, respectively, and $\tau^{0}_{t}, \tau^{1}_{t}, ...$ are the indices of the reference slices for the $t$-th slice, i.e., the slices on which $\hat{z}_t$ is conditioned, such that $\hat{z}_{\tau^{0}_{t}}, \hat{z}_{\tau^{1}_{t}}, ...$ are at the higher hierarchical levels and are always coded prior to $\hat{z}_t$. By introducing the in-parameter coding structure, each slice depends on the slices at the previous levels but not on all other slices within the same grid tensor. Different from the common approach in neural codecs, which applies the coding structure across temporal frames, where each frame can be individually decoded into the corresponding video frame, our \textit{in-parameter coding structure} approach utilizes slices in the feature grids, each of which can be shared across multiple video frames due to the use of temporal interpolation \cite{lee2023ffnerv, kwan2024hinerv, kim2023c3}. We apply the same process to different resolution grid parameters.

\vspace{5pt}\noindent\textbf{Practical implementation for optimization.}  Although structures such as \textit{low-delay} or \textit{hierarchical B-frame} styles have been widely adopted for various neural video codecs \cite{lu2019dvc, li2021deep}, their common implementations are not compatible with overfitted codecs, especially for long video sequences. In particular, most neural video codecs utilize quantization step sizes and offsets\footnote{In related works, the mean of the distributions (e.g. Gaussian) is commonly implemented as the quantization offset in their actual implementations.}, which are conditioned on the previously coded frames. This creates a \textit{feature propagation chain} that accumulates over video frames. Directly deploying this approach for overfitting long sequences is inefficient or even impossible; the network activations for all frames must be cached to compute gradients during backpropagation, requiring memory far beyond what current GPUs can provide.

To address this issue, we propose two modifications in NVRC++ when deploying the coding structure: (i) NVRC++ utilizes quantization with step sizes that are not conditioned on previous frames, and eliminates the use of offsets; and (ii) due to the removal of the feature propagation chain, NVRC++  can perform random sampling during the optimization process, which reduces the memory footprint while maintaining training-testing consistency. By utilizing this unconditional and symmetric quantization, NVRC++ not only improves the efficiency of rate estimation but also decouples it from distortion computation. For instance, rate and distortion can be calculated separately to reduce both memory footprint and computational cost, following \cite{kwan2024nvrc}. \cref{fig:coding_structure} (Right) shows a comparison between the conventional and our proposed optimization pipelines, where we utilize random sampling during rate estimation and skip the entropy models in distortion computation.

During rate estimation, we scale the computed rate using its expectation to ensure consistency when balancing the rate-distortion trade-off. As a result, our largest model, which contains feature grids with over 100M parameters, can still encode a 600-frame 1920$\times$1080 video sequence \cite{mercat2020uvg}, when running on a consumer GPU with only 24GB of memory.

\vspace{5pt}\noindent\textbf{Feature grid masking.} Finally, we observe that when incorporating high-resolution features, overfitted codecs utilizing multi-resolution feature grids can suffer from reduced performance at low bitrates. This is likely due to the multi-resolution grids, where output pixel features are obtained via interpolation. In lower-resolution grids, multiple neighboring pixels share the same underlying features because the output resolution is significantly higher than the grid resolution; this allows the model to effectively exploit spatio-temporal redundancies. However, high-resolution grids can easily fit video details, allowing them to assign features to very small regions or even individual pixels. Consequently, these high-resolution features are shared less effectively across the sequence, preventing the effective exploitation of spatio-temporal redundancy at lower bitrates.

To address this issue, we propose a random masking mechanism, following the concept of Dropout \cite{hinton2012improving}, for feature grids with higher resolutions at the beginning of training. This forces the model to utilize lower-level grids during the early stages of optimization. The dropout is applied for both rate and distortion calculations, ensuring that the balance between the two terms remains intact. The dropout ratio is annealed throughout the training process, eventually reaching zero in the later stages. We apply different schedules to different sets of grid features depending on their resolution (details are provided in  \textit{Supplementary}). We found that this simple approach effectively balances performance across different bitrates, as shown in the experimental results.

\subsection{Multi-dimensional grid entropy model}
\label{subsec:gridcoding}

Although higher dimensional grids allow INR models to achieve improved reconstruction quality without increasing complexity, they are costly in storage, limiting their practical applications in video coding. Recent work utilizing an autoregressive model \cite{minnen18joint} for entropy modeling \cite{kim2023c3, kwan2024nvrc, jia2025towards}, while demonstrating promising performance, is impractical due to its slow, sequential coding process. Here, we introduce an improved grid entropy model, inspired by the advances in autoencoder-based image and video compression \cite{balle2018variational, he2021checkerboard, li2021deep, li2022hybrid, li2023neural}, where different priors are used to enhance entropy modeling while providing high coding speed.

As shown in \cref{fig:main}, our grid entropy model leverages (i) scale priors, where larger scale feature grids are conditioned by grids with a lower dimension in a hierarchical manner \cite{balle2018variational}; (ii) temporal priors, which we code grid slices sequentially and utilize the coded slices as conditions \cite{li2021deep}, as mentioned in \cref{subsec:overfitting}; and (iii) spatial priors, following common approaches in neural coding to utilize spatial context to improve compression \cite{he2021checkerboard, li2022hybrid, li2023neural}.

For a set of multi-scale feature grid parameters $\{ z^{l} \mid 0 \le l < L_{grid} \}$, where $L_{grid}$ is the number of grid levels, we
first encode the grids at a lower resolution, as they will be utilized as the condition for coding higher resolution grids. For coding a particular quantized feature slice $\hat{z}^{l}_{t}$, we gather the reference slices in the same scale, e.g., $\hat{z}^{l}_{\tau^{{0}}_{t}}$, as the temporal prior, and the coarser level slice, $\hat{z}^{l - 1}_{t}$, as the condition. Here, we obtain $\hat{z}^{l - 1}_{t}$ by linear interpolation when $\hat{z}^{l - 1}$ has a lower resolution than $\hat{z}^{l}$. Finally, we perform the coding of $\hat{z}^{l}_{t}$ in a multi-step manner, following the spatial prior used for image and video coding, where we mask the grid slice $\hat{z}^{l}_{t}$ by $m_{s}(\hat{z}^{l}_{t})$ at the $s$-th step and utilize the previously coded features in the same slice, $m_{0}(\hat{z}^{l}_{t})$, ..., $m_{s-1}(\hat{z}^{l}_{t})$, as the condition, where $m_{s}(\cdot)$ is the masking pattern at the $s$-th step following \cite{he2021checkerboard, li2022hybrid, li2023neural}.

While using a larger number of coding steps can improve performance, we utilize 1-4 coding steps, depending on the resolution, to reduce the amount of computation and memory footprint. The entropy model first employs the quantization level with a specific step size $\delta^{l}$, which is shared between all temporal and spatial positions to perform quantization: $\hat{z}^{l}_{t} = \lfloor \frac{z^{l}_{t}}{\delta^{l}}\rceil \cdot \delta^{l}$. The distribution of entropy parameters, including the mean $\mu^{l, t, s}$ and scale $\sigma^{l, t, s}$,  will then be estimated by:
\begin{equation}
    p(m_{s}(\hat{z}^{l}_{t})|\hat{z}^{l-1}_{t}, \hat{z}^{l}_{\tau^{0}_{t}}, \dots,\dots, m_{0}(\hat{z}^{l}_{t}), \dots).
\end{equation}
This implementation details can be found in \textit{Supplementary}.

\subsection{Implicit Neural Representation}
\label{subsec:HiNeRV++}

In NVRC++, we developed an efficient neural representation for video coding, named HiNeRV++ (shown in \cref{fig:main}, with a full-scale figure provided in \textit{Supplementary}). HiNeRV++ is an enhanced variant of HiNeRV \cite{kwan2024hinerv},  but with the following modifications.

\vspace{5pt}\noindent\textbf{Higher resolution grid features.} Since one of our focuses is to improve coding scalability with respect to the rate and quality levels, we utilize multiple feature grids for feature learning, where the grid resolution is much higher than that of HiNeRV. In HiNeRV, there are four stages of network blocks, where the blocks in each stage perform the computation in the same resolution. Here we utilize encodings extracted from different sets of feature grids \cite{kwan2024hinerv} as the input to the first three stages of the network blocks, while HiNeRV only does this in the first stage. We also continue using multi-resolution grids in the first input stage.

\vspace{5pt}\noindent\textbf{Multi-reference inputs.} In the HiNeRV variant used in NVRC \cite{kwan2024nvrc}, 3D convolution is employed during the encoding process to extract the input encoding from the feature grids. To provide the INR with stronger input conditions, we propose utilizing multiple feature slices as the input for HiNeRV++, as shown in \cref{fig:coding_structure} (Left). Specifically, we gather both the interpolated features and multiple neighboring feature grid slices in the original grid to serve as the input. This provides the model with higher-quality input encodings, as the raw, nearby slices are not subject to interpolation. Consequently, the model can learn to directly leverage these high-fidelity encodings rather than solely relying on interpolated features, as in the original HiNeRV \cite{kwan2024hinerv}. We apply this modification to all encoding layers.

\vspace{5pt}\noindent\textbf{Removal of the highest level blocks.} The blocks in the last stage, together with the head layer in HiNeRV++, are replaced by a single pixel shuffle layer \cite{shi2016real}. Although pixel shuffle layers have high parameter complexity with the number of channels and are not efficient for the INR-based compression task \cite{kwan2024hinerv}, they are efficient when used as the output layer, since there are only a few channels in the pixel domain. Furthermore, the computation and memory costs in the highest resolution are substantial, which could slow down inference if stronger blocks, such as ConvNeXt blocks \cite{liu2022conv}, are placed at that resolution.

\section{Experimental Settings}
\label{sec:experiment}

\noindent\textbf{Test datasets.} The evaluation of the NVRC++ framework is based on commonly used datasets: UVG \cite{mercat2020uvg} and MCL-JCV \cite{wang2016mcl}. Both of them contain full HD ($1920\times1080$) sequences, where the frame counts for each sequence range from 300 to 600 for UVG and from 120 to 150 for MCL-JCV. These datasets comprise 7 and 30 sequences, respectively.

\vspace{5pt}\noindent\textbf{Implementation details.} NVRC++ is implemented based on HiNeRV \cite{kwan2024hinerv} and NVRC \cite{kwan2024nvrc}, with the same training strategy. Unlike HiNeRV and NVRC, we configured NVRC++ with four different complexity scales (S1 (smallest), S2, S3 and S4 (largest)), by varying the decoder and local grid channels in HiNeRV++. Each scale employs a single model configuration that achieves a wide quality range through the use of high-resolution feature grids and the advanced entropy model. This allows us to evaluate the trade-off between decoding complexity and compression performance across different computational budgets. More details on implementation are provided in \textit{Supplementary}.

\vspace{5pt}\noindent\textbf{Benchmark methods.} For benchmarking NVRC++, three conventional codecs, including x265 (\textit{veryslow}) \cite{x265}, HM-18.0 (\textit{randomaccess}) \cite{hm}, and VTM-20.0 (\textit{randomaccess}), are employed. Two autoencoder-based neural video codecs, DCVC-FM \cite{li2024neural} and DCVC-RT \cite{jia2025towards}, are included, both using the single intra-frame setting for the best performance. Additionally, three state-of-the-art INR-based codecs, HiNeRV \cite{kwan2024hinerv}, NVRC \cite{kwan2024nvrc} and C3 \cite{kim2023c3} have also been included in this experiment. All results are obtained from the original paper or produced by their open-source implementations and the checkpoints, if provided.

\vspace{5pt}\noindent\textbf{Evaluation metrics.} The evaluation was performed in the RGB color space with the BT.601 color conversion. PSNR and MS-SSIM are used here to assess video quality, based on which Bj{\o}ntegaard Delta rate (BD-rate) \cite{bdrate} figures are calculated for NVRC++ and each benchmark codec against x265 (\textit{veryslow}). Full sequences are used in the evaluation.

\vspace{5pt}\noindent\textbf{Additional results.} We provide additional experimental results, including evaluations on the JVET-CTC Class B dataset and in the YUV420 color space, in the \textit{Supplementary}.

\begin{table}[t!]
    \scriptsize
    \centering
    \caption{BD-rate results of the proposed NVRC++ model (configured at four scales S1-S4) against conventional and neural video codecs on UVG and MCL-JCV datasets. Coding complexity is reported in terms of kMACs per pixel and Frames Per Second (FPS). 
    HiNeRV and NVRC complexities are presented as ranges across various quality scales. FPS are obtained using the same platform with the NVIDIA RTX 4090 GPU with FP16 mixed precision. For NVRC/NVRC++/C3, we report the synthesis transform and the entropy coding time separately following \cite{kwan2024nvrc}. *: with custom CUDA kernel. **: Entropy coding is not implemented in the official code.}
    \resizebox{1\linewidth}{!}{

    \begin{tabular}{lccccccc}
        \toprule
        \multicolumn{1}{r}{BD-rate (\%)}& \multicolumn{2}{c}{UVG} & \multicolumn{2}{c}{MCL-JCV} & \multicolumn{1}{c}{Enc. Complexity} & \multicolumn{2}{c}{Dec. Complexity} \\
        \cmidrule(lr){2-3}  \cmidrule(lr){4-5} \cmidrule(lr){6-6} \cmidrule(lr){7-8}
        Method & PSNR & MS-SSIM & PSNR & MS-SSIM & FPS & kMACs/Pixel & FPS \\
        \midrule
        x265 (\textit{veryslow}) \cite{x265} & 0.00 & 0.00 & 0.00 & 0.00 & - & - & - \\
        HM (\textit{RA}) \cite{hm}           & -44.54 & -43.85 & -39.91 & -41.61 & - & - & - \\
        VTM (\textit{RA}) \cite{vtm}         & -62.81 & -61.74 & -58.86 & -61.66 & - & - & - \\
        \midrule
        DCVC-FM \cite{li2024neural}          & -66.30 & -67.16 & -58.33 & -59.45 & 5.9 & 369.6 (I)/865.5 (P) & 3.7 \\
        DCVC-RT \cite{jia2025towards}        & -65.56 & -68.05 & -57.40 & -61.69 & 113.0 & 364.9 (I)/166.8 (P) & 57.8/141.9* \\
        \midrule
        HiNeRV \cite{kwan2024hinerv}         & -45.84 & -63.04 & -28.75 & -44.79 & 0.014-0.026 & 87.3-346.3 & 13.3-49.9 \\
        NVRC \cite{kwan2024nvrc}             & -73.74 & -80.65 & -51.61 & -66.83 & 0.006-0.016 (13.8-26.2) & 173.4-930.6 & 14.0-33.2 (5.6-19.5) \\
        C3 \cite{kim2023c3}                  & -21.91 & -14.06 & -19.42 & -38.35 & 0.0004 & 4.4 & 434.0 (N/A**) \\
        \midrule
        \textbf{NVRC++ (S1)}                 & -40.06 & -67.59 & -22.07 & -58.02 & 0.020 (74.9) & 7.3 & 365.0 (62.0) \\
        \textbf{NVRC++ (S2)}                 & -61.41 & -76.78 & -47.61 & -67.80 & 0.015 (75.0) & 25.1 & 162.9 (62.4) \\
        \textbf{NVRC++ (S3)}                 & -71.20 & -81.25 & -55.90 & -70.80 & 0.010 (75.9) & 92.5 & 73.4 (62.8) \\
        \textbf{NVRC++ (S4)}                 & -74.82 & -81.88 & -51.84 & -67.30 & 0.005 (70.5) & 357.7 & 34.0 (59.23) \\
        \bottomrule
    \end{tabular}
    }
    \label{tab:bdrate_comprehensive}
    \vspace{-5pt}
\end{table}

\section{Results and Discussion}
\label{sec:results}

\subsection{Overall Performance}
\begin{figure}[!t]
    \centering
    \begin{subfigure}[t]{0.64\linewidth}
     \vspace{0pt}
        \centering
        \includegraphics[width=0.49\linewidth]{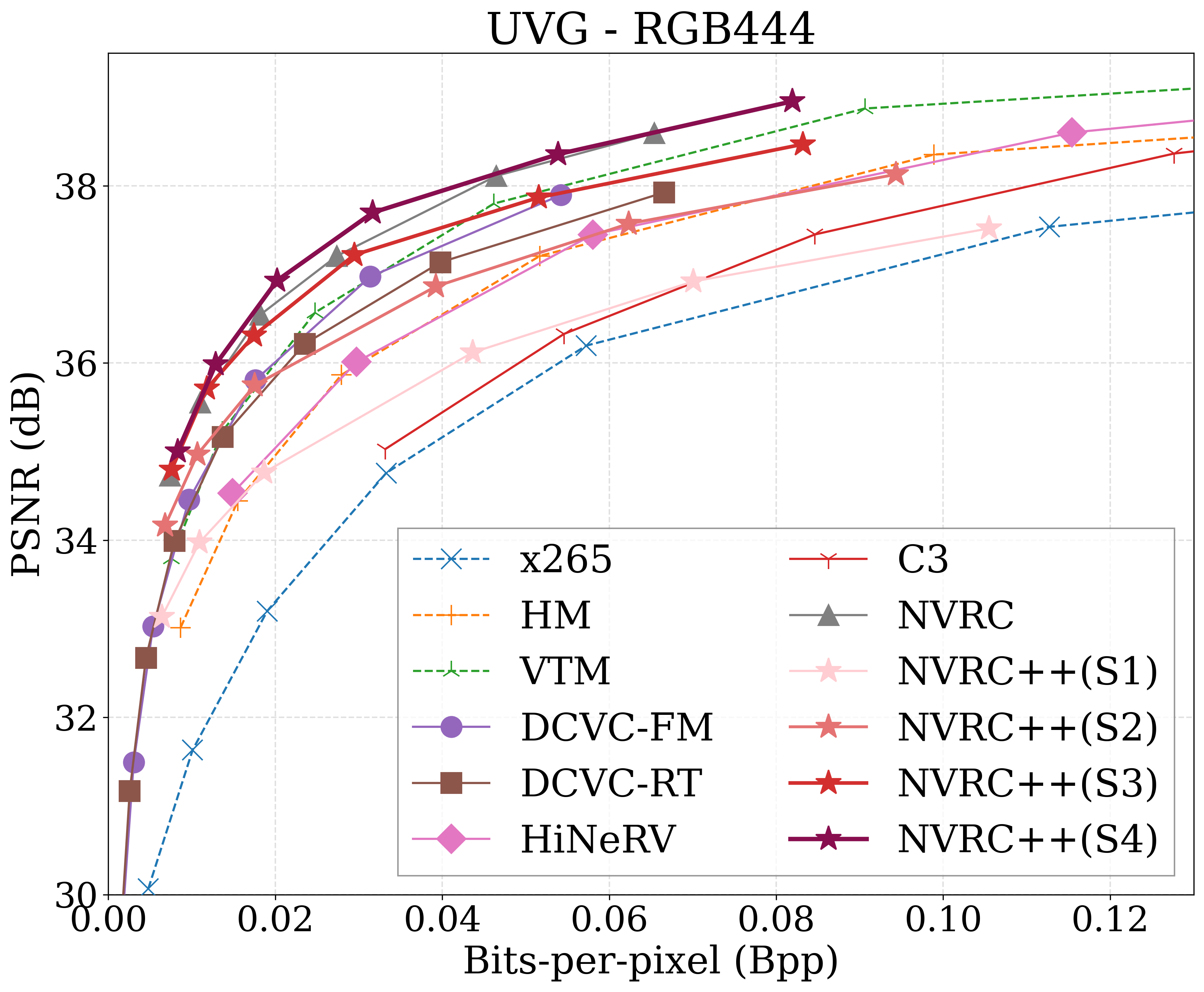} \hfill
        \includegraphics[width=0.49\linewidth]{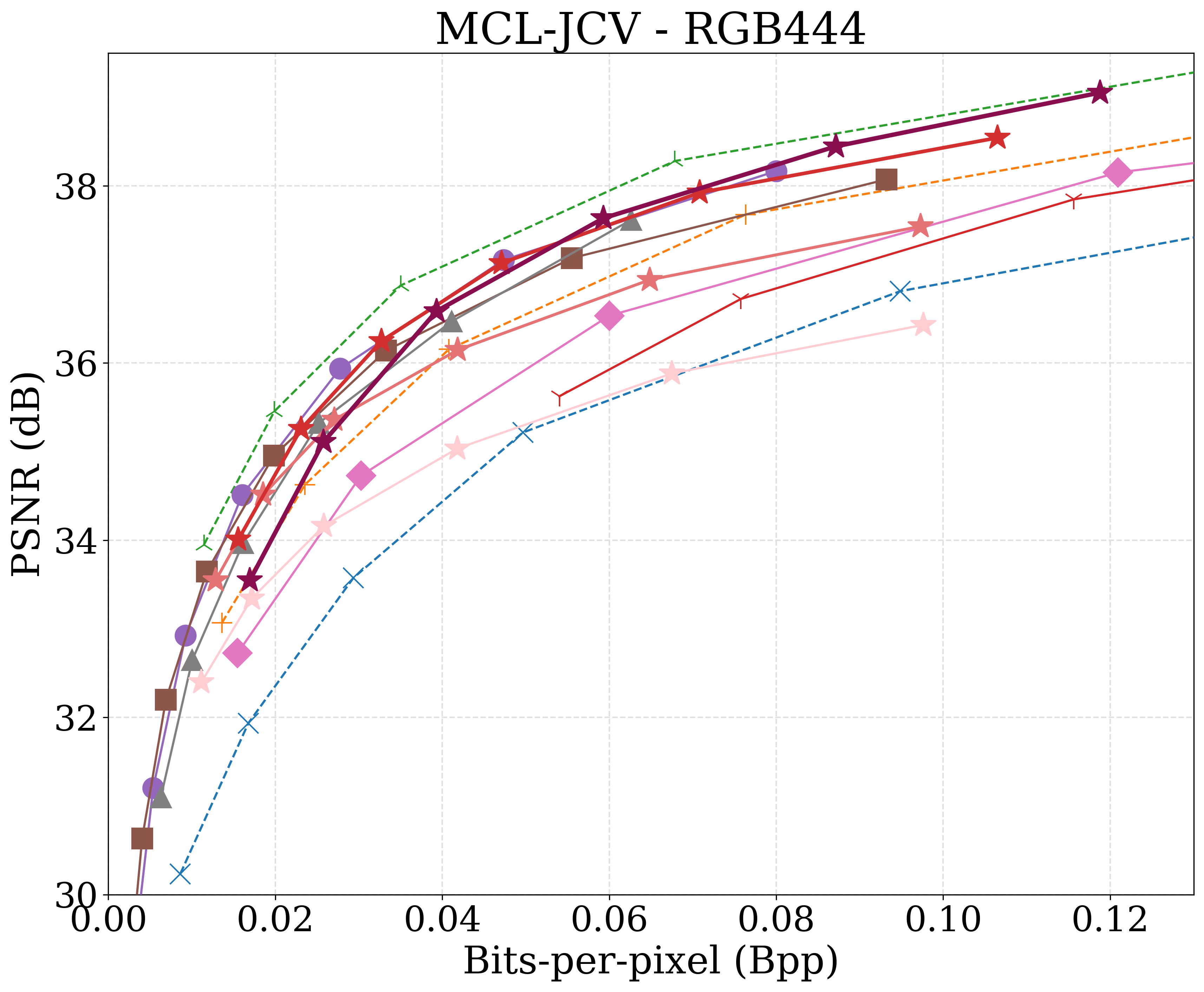}

        \includegraphics[width=0.49\linewidth]{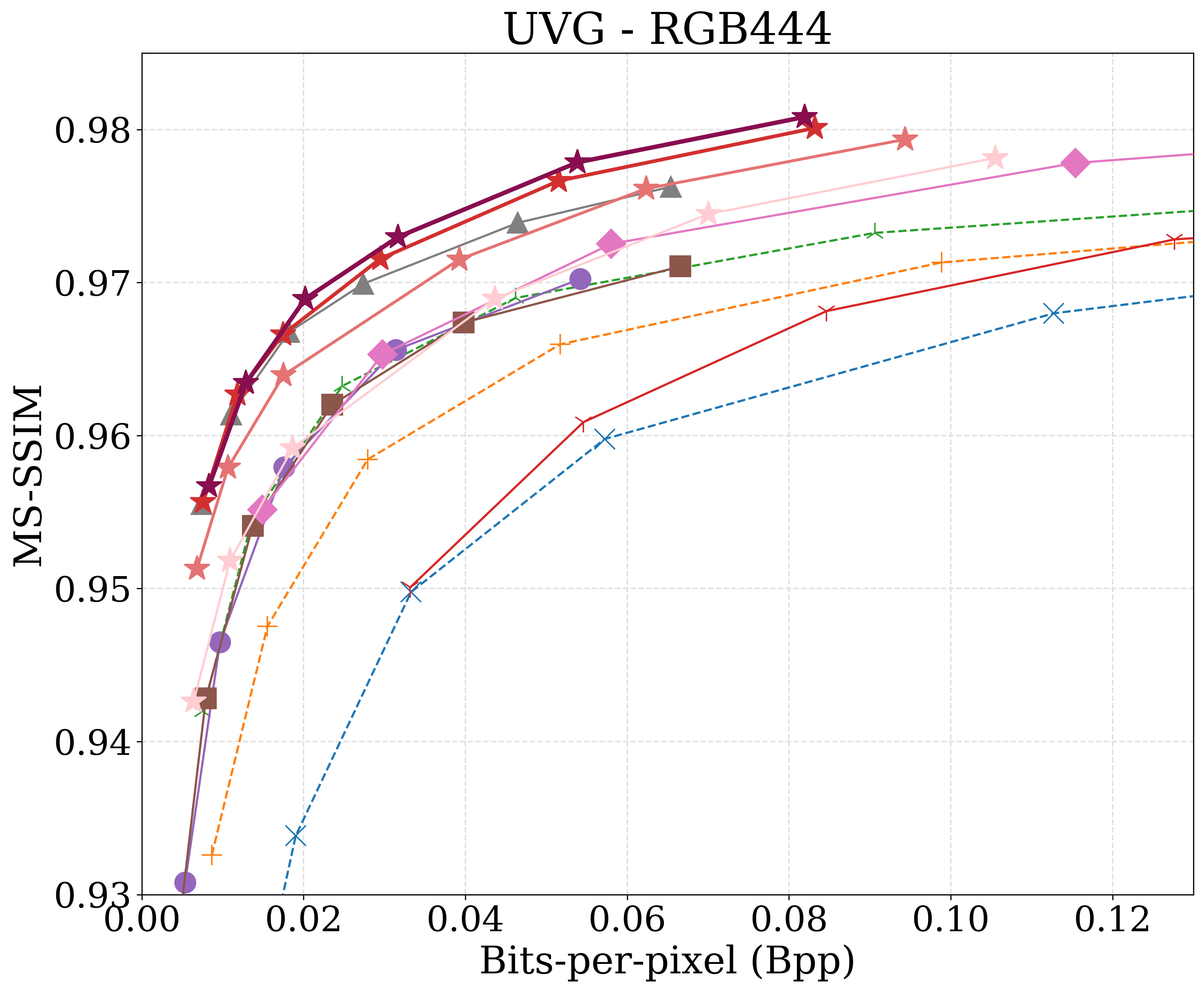} \hfill
        \includegraphics[width=0.49\linewidth]{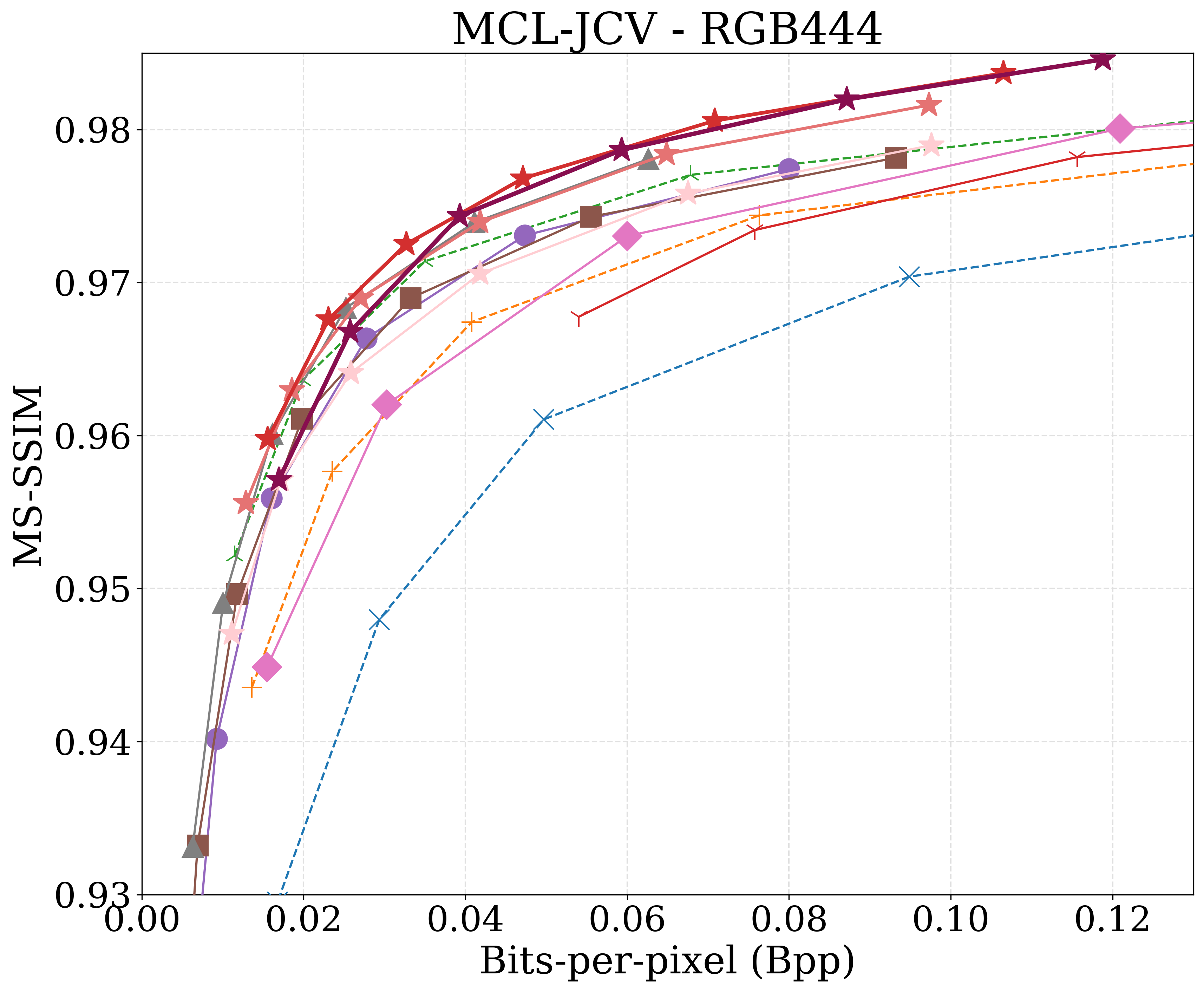}
           \vspace{-15pt} 
    \end{subfigure}
    \hfill
    \begin{subfigure}[t]{0.32\linewidth}
     \vspace{0pt}
        \centering
        \includegraphics[width=1\linewidth]{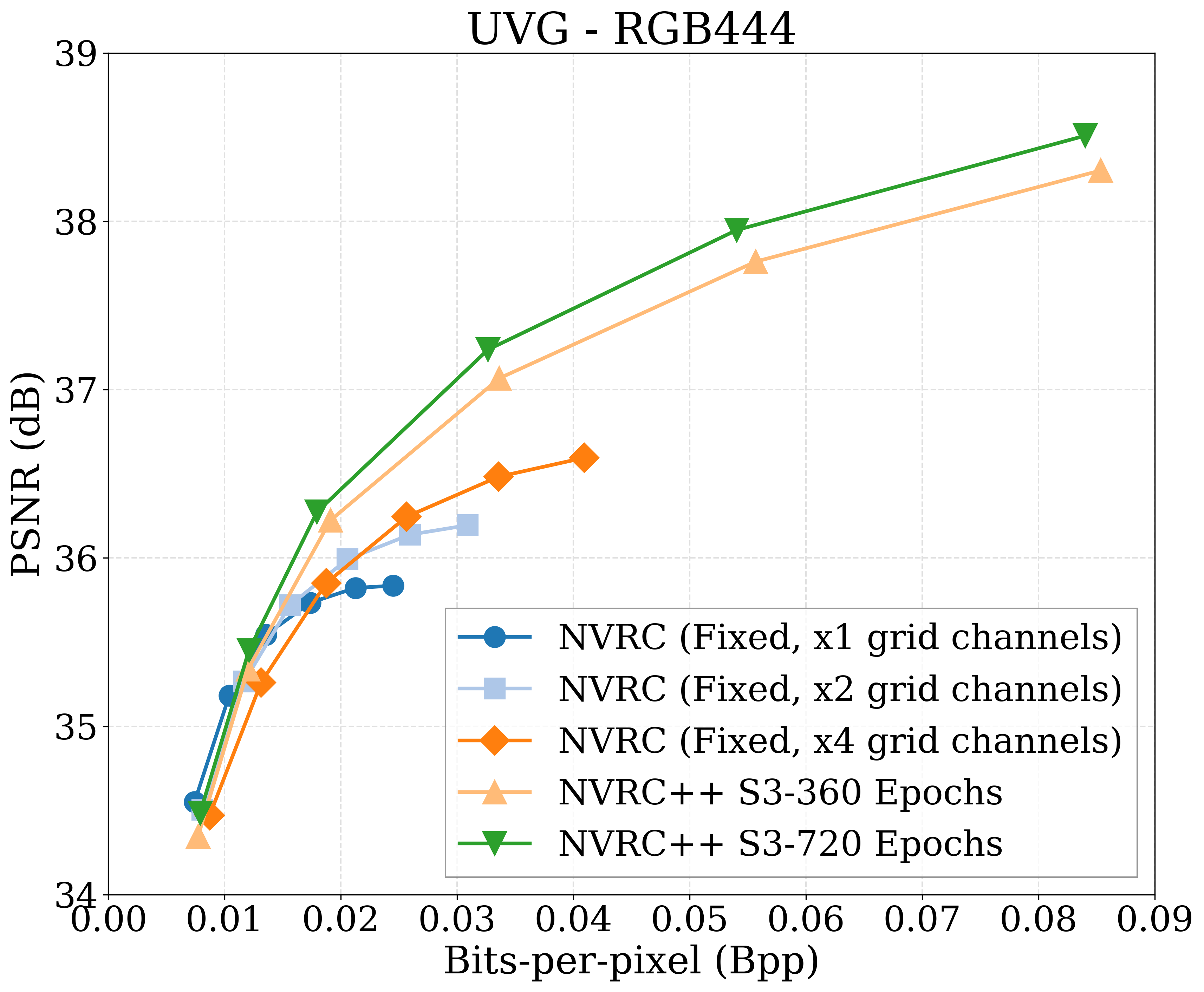}

        \includegraphics[width=1\linewidth]{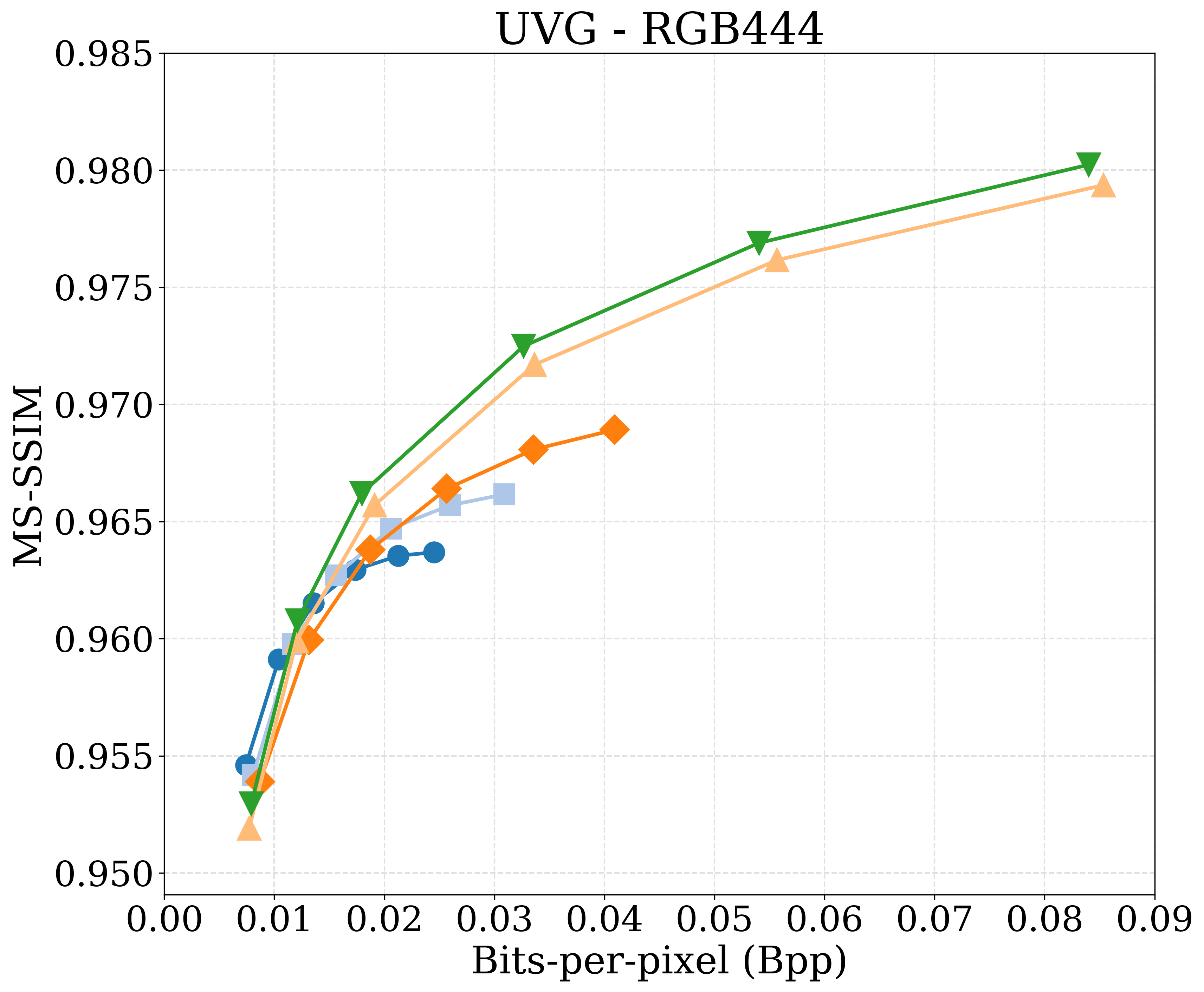}
           \vspace{-15pt} 
    \end{subfigure}
    \caption{(Left) RD curves of benchmarked codecs. (Right) Comparison of NVRC++ (92.5kMACs/px) with NVRC (with fixed complexity (102.7kMACs/px)) for scalability. }
    \label{fig:rdcurve}
   \vspace{-15pt} 
\end{figure}

The overall performance of the proposed NVRC++ framework is summarized in \cref{tab:bdrate_comprehensive}. The corresponding rate-quality curves are shown in \cref{fig:rdcurve} (Left).

\vspace{5pt}\noindent\textbf{Compression performance across scales.}
NVRC++ demonstrates highly scalable compression across four variants (S1-S4), progressively matching or exceeding the baselines of x265, HM, VTM, and the SOTA NVRC. It excels particularly on long sequences (e.g., UVG dataset). At the highest capacity, S4/S3 outperform NVRC's performance in UVG/MCL-JCV while decoding up to 5.4x/8.5x faster. In the mid-complexity range, S3 outperforms VTM and DCVC-RT, and the lightweight S2 beats HM and rivals DCVC-RT at low bitrates, achieving DCVC-level quality (in MS-SSIM) at a fraction of their computational cost. Finally, the ultra-lightweight S1 comfortably outperforms x265 and recent INRs like C3 at similar complexities, while rivaling HiNeRV and HM in MS-SSIM.

\vspace{5pt}\noindent\textbf{Decoding speed and complexity.} A key advantage of NVRC++ is its constant decoding complexity across bitrates for a given scale. Unlike NVRC \cite{kwan2024nvrc} and HiNeRV \cite{kwan2024hinerv}, which exhibit bitrate-dependent costs (173--931 and 87--346 kMACs/Pixel), Each NVRC++ variant maintains fixed complexity for different rate/quality points. It is up to 8.5 $\times$ faster than NVRC in decoding speed (full decoding) . The overall complexity-performance trade off of NVRC++ and other benchmark codecs is illustrated by \cref{fig:tradeoff}. 

\begin{figure}[t!]
    \centering
    \scriptsize
    \begin{subfigure}[b]{0.99\linewidth}  
        \centering
        \includegraphics[width=\linewidth]{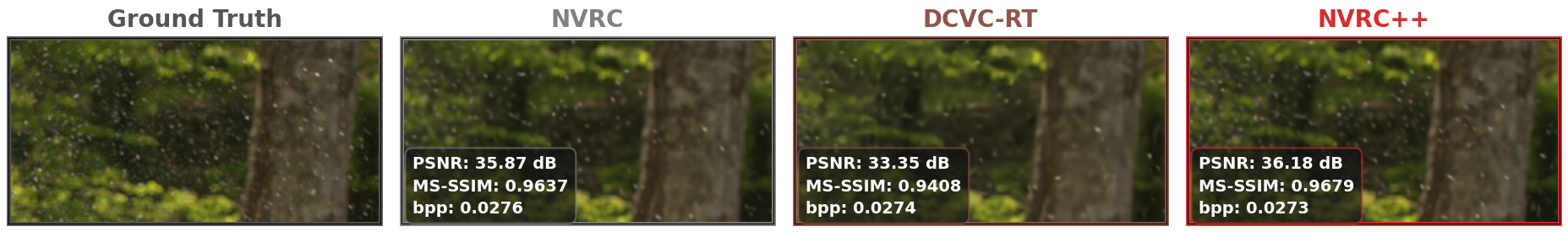}
    \end{subfigure}
    \begin{subfigure}[b]{0.99\linewidth}  
        \centering
        \includegraphics[width=\linewidth]{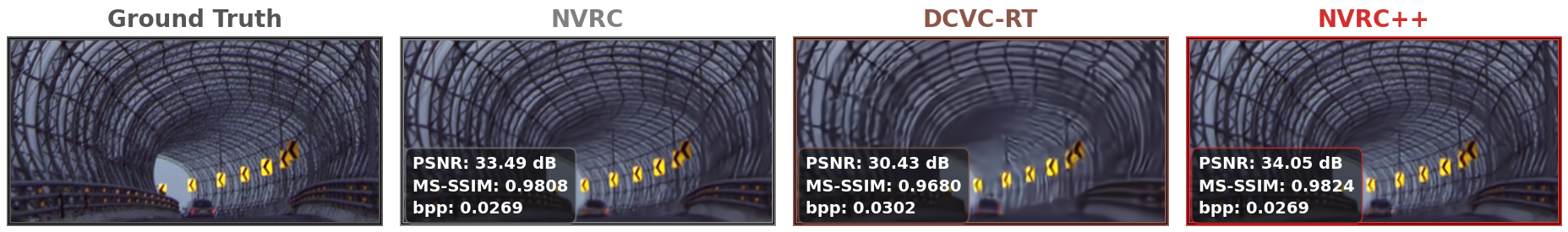}
    \end{subfigure}
    \caption{Qualitative comparison between GT, NVRC, DCVC-RT and NVRC++ (ours).}
    \label{fig:visualcomparision}
\vspace{-15pt}
\end{figure}

\vspace{5pt}\noindent{\textbf{{Qualitative results.}}} We provided a visual comparison between the proposed method and two SOTA neural video codecs, NVRC and DCVC-RT in \cref{fig:visualcomparision}. It can be observed that, with similar or even lower bitrates, NVRC++ produces reconstructed content with improved visual quality compared to these two benchmarks, which further confirms its coding performance in perceptual quality.

\subsection{Ablation study}
The effectiveness of primary innovations in NVRC++ has been validated in an ablation study based on the UVG dataset. The results are summarized in \cref{tab:ablation}. Specifically, we replaced the multiple high resolution feature grids with only the low-resolution grids \cite{kwan2024hinerv} (v1), removed the random feature grid masking strategy (v2), disabled the spatial prior (v3), temporal prior (v4), and scale prior (v5) in the entropy model, substituted our multi-reference input design with a 2D convolutional stem (v6) and a 3D convolutional stem (v7). It can be observed that all these test variants are associated with inferior performance compared to the original NVRC++ - this confirms the contribution of each component tested. A more detailed analysis is provided in \textit{Supplementary}.

\subsection{Additional analysis}
\noindent\textbf{HiNeRV++ vs HiNeRV.} We have also validated the proposed HiNeRV++ architecture by comparing it with the original HiNeRV, with results shown in \cref{tab:ablation2}. It is noted that by offloading the representational burden from neural layers to high-resolution feature grids, HiNeRV++ drastically increases grid parameters while reducing layer complexity, leading to a lower computational cost and a substantial increase in decoding speed. Furthermore, HiNeRV++ exhibits faster convergence and superior final quality, outperforming HiNeRV by 2.09dB in PSNR at 300 epochs with just 37 epochs of training, proving that lightweight networks with high-resolution grids are a superior solution for INR-based video compression.

\vspace{5pt}\noindent\textbf{Scalability of NVRC++.} 
We also performed a comparison between NVRC and NVRC++ regarding their scalability with respect to quality/rate. For NVRC, we re-implemented the model with a fixed complexity using varying numbers of feature grid channels. For NVRC++, we utilized the S3 variant. Our results in \cref{fig:rdcurve} (Right) show that NVRC achieves a range of quality levels by varying the number of grid channels, at the cost of reduced compression performance at lower bitrates. In contrast, NVRC++ achieves a significantly wider quality range while maintaining high compression performance across all bitrates. This improvement is attributed to both the use of higher-resolution grids and the proposed optimization mechanisms.

\begin{table}[!t]
    \centering

    \begin{minipage}[t]{0.44\textwidth}
        \centering

        \caption{The ablation study results on the UVG dataset. Here NVRC++ (S2) is used as the anchor. }
   \vspace{-5pt} 
        \resizebox{1\linewidth}{!}{
            \begin{tabular}{lrr}
                \toprule
                BD-rate (\%) & PSNR & MS-SSIM \\
                \midrule
                (v1) w/ Low resolution grid & 44.01 & 59.00 \\
                (v2) w/o Masking & 30.33 & 39.77 \\
                (v3) w/o Spatial prior & 1.42 & 0.52 \\
                (v4) w/o Temporal prior & 5.13 & 4.00 \\
                (v5) w/o Scale prior & 6.12 & 6.60 \\
                (v6) w/ 2D stem & 35.55 & 31.64 \\
                (v7) w/ 3D stem & 7.76 & 8.44 \\
                \bottomrule
            \end{tabular}
        }
        \label{tab:ablation}
    \end{minipage}
    \hfill
    \begin{minipage}[t]{0.51\textwidth}
        \centering
        \caption{Comparison between HiNeRV (S) and HiNeRV++ (S2). Results are measured on the UVG dataset.}
   \vspace{-5pt} 
        \resizebox{\linewidth}{!}{
            \begin{tabular}{llcc}
                \toprule
                \multicolumn{2}{c}{} & \textbf{HiNeRV} & \textbf{HiNeRV++} \\
                \midrule
                Parameters & Grids/Layers & \textbf{0.35M}/2.84M & 100.06M/\textbf{1.23M} \\
                \midrule
                \multirow{2}{*}{Complexity} & kMAC/px & 87.3 & \textbf{24.8}\\
                                            & Dec. FPS & 35.5 & \textbf{162.9} \\
                \midrule
                \multirow{4}{*}{PSNR (dB)}       & @37 Epochs & 33.70 & \textbf{37.36} \\
                                            & @75 Epochs & 34.53 & \textbf{38.27} \\
                                            & @150 Epochs & 34.99 & \textbf{39.03} \\
                                            & @300 Epochs & 35.27 & \textbf{39.46} \\
                \bottomrule
            \end{tabular}
        }
        \label{tab:ablation2}
    \end{minipage}
   \vspace{-5pt} 
\end{table}

\section{Conclusion}
\label{sec:conclusion}

In this paper, we have proposed an enhanced neural video representation compression framework, NVRC++, aiming to address the low scalability issue associated with existing INR-based video codecs. Through integrating new innovations, including multiple high-resolution feature grids, a multi-dimensional grid entropy model, and an in-parameter coding structure, NVRC++ has significantly improved its scalability, enabling the use of a single model configuration for a wide range of bitrates and quality levels. More importantly, it is associated with a higher encoding efficiency (up to 3.3 times faster than NVRC) and achieves real-time decoding (73.4 fps for frame decoding), while offering similar coding performance to NVRC. This work represents an important step forward towards practical applications of INR-based video coding methods.

\section*{Acknowledgements}
This work was supported by UK EPSRC (iCASE Awards), BT, the UKRI MyWorld Strength in Places Programme and the University of Bristol. Part of the computational work was also supported by the facilities provided by the Advanced Computing Research Centre at the University of Bristol. The authors acknowledge the use of resources provided by the Isambard-AI National AI Research Resource (AIRR) \cite{mcintoshsmith2024isambard}. Isambard-AI is operated by the University of Bristol and is funded by the UK Government’s Department for Science, Innovation and Technology (DSIT) via UK Research and Innovation; and the Science and Technology Facilities Council [ST/AIRR/I-A-I/1023].

\bibliographystyle{splncs04}
\bibliography{main}

\clearpage

\renewcommand{\theequation}{A\arabic{equation}}
\renewcommand{\thefigure}{A\arabic{figure}}
\renewcommand{\thetable}{A\arabic{table}}
\renewcommand{\thesection}{A\arabic{section}}
\setcounter{equation}{0}
\setcounter{figure}{0}
\setcounter{table}{0}
\setcounter{section}{0}

\begin{center}
    {\large\bfseries Enhanced Neural Video Representation Compression Across Extreme Complexity and Quality Scales}\\[1.5ex]
    {\large\itshape Supplementary Material}
\end{center}

\vspace{2ex}

\section{Overview}
\label{sec:supp_overview}
This supplementary document provides additional technical details, extended experimental results, results discussion, and qualitative visualizations to support the main paper. The implementation will be made available at the
\href{https://hmkx.github.io/nvrc-pp/}{project page}.

\section{Architecture details}
\label{sec:arch_details}

\subsection{Overall architecture}
\label{subsec:overall_architecture}

In the proposed NVRC++, we build upon the architecture of NVRC \cite{kwan2024nvrc}, utilizing an implicit neural representation (INR) paired with a hierarchy of entropy models. These models are designed to compress various groups of INR parameters, their corresponding entropy model parameters, and the associated quantization parameters. 

This hierarchical framework incorporates our proposed grid entropy model, which leverages multiple priors for coding feature grids (\cref{subsec:entropy_model}), alongside the HiNeRV++ architecture (\cref{subsec:inr}). Furthermore, following the original NVRC \cite{kwan2024nvrc}, we employ a dual-axis Gaussian model for INR weight compression (the weight entropy model) and a conditional Gaussian model to compress the entropy and quantization parameters of both the grid and weight models (the meta entropy models). Finally, the highest-level parameters within the unconditional entropy model are stored in \texttt{FP16} format, which introduces only a negligible storage overhead.

\subsection{Entropy model architecture}
\label{subsec:entropy_model}

 \begin{figure}[!t]
     \centering
     \includegraphics[width=0.5\linewidth]{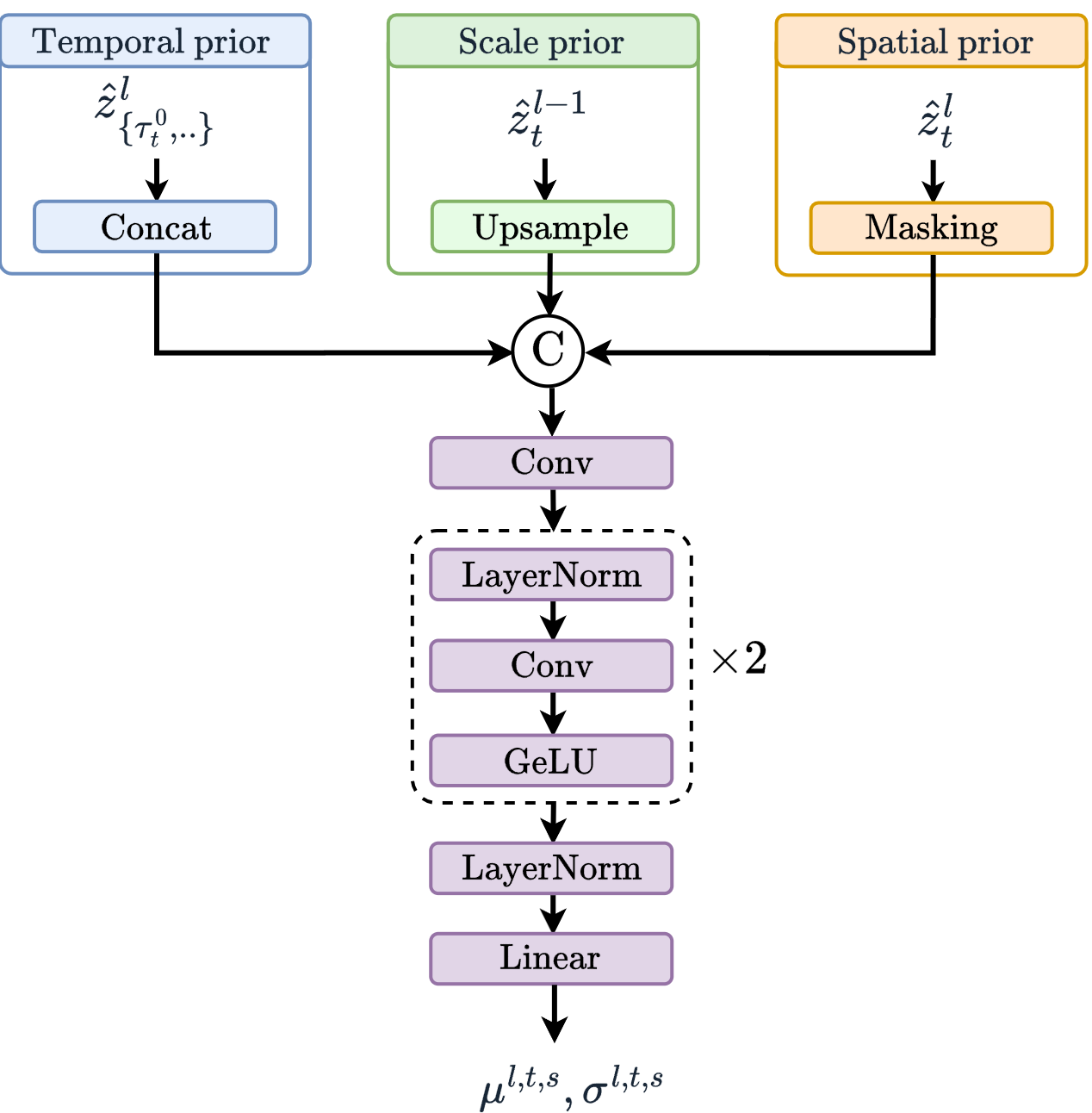}
     \caption{\textbf{Architecture of the grid entropy model.} Note that, we utilize distinct parameter sets at different coding steps $s$ (i.e., under varying spatial masking conditions) for the first convolutional layers. The entropy model predicts the mean $\mu^{l,t,s}$ and log-scale $\sigma^{l,t,s}$ for the $l$-th level feature grid at each temporal slice $t$.}
     \label{fig:grid-em}
 \end{figure}

As discussed in the main paper, we utilize a multi-prior entropy model to enhance the compression of the high-dimensional feature grids introduced by the INR framework (\cref{subsec:inr}). Inspired by recent advances in neural compression \cite{balle2018variational, he2021checkerboard, li2021deep, li2022hybrid, li2023neural}, the proposed model incorporates scale, temporal, and spatial priors for feature encoding. This approach significantly accelerates both entropy encoding and decoding compared to the original NVRC \cite{kwan2024nvrc} and related works, which rely on autoregressive models that offer high coding performance but suffer from slow practical execution speeds.

Our entropy model treats each feature grid as a stack of temporal slices, coding them in a hierarchical manner. To estimate the distribution parameters (mean and scale) of a particular slice, the model concatenates inputs from multiple contexts: the corresponding slice at a lower scale (resolution) - scale prior, previously coded temporal slices following a Hierarchical-B coding structure - temporal prior, and the decoded context from the current slice - spatial prior. 

The network architecture comprises a stack of convolutional layers interleaved with Layer Normalization \cite{ba2016layer} and GELU activations \cite{hendrycks2016gaussian}. This is followed by an output linear layer that predicts the parameters (i.e., the mean and log-scale) of the estimated Gaussian distribution. We apply a log-scale offset of -4 to obtain a more concentrated initial distribution, inspired by \cite{kim2023c3}.

We maintain a fixed number of entropy model layers across different configurations, scaling the network width dynamically based on the number of feature grid channels. Specifically, the hidden channel dimension is strictly set to 8 times the number of input feature grid channels. 

To encode temporal slices, we employ a Hierarchical-B coding structure spanning up to 10 layers. The number of slices encoded at each progressive layer follows the sequence $1, 1, 2, 4, \dots, 128$, respectively. To accommodate long video sequences which the feature grids could contain up to 600 slices, any remaining slices are cascaded into the final layer. Furthermore, we utilize up to four previously coded reference slices for the temporal prior.

\subsection{INR architecture}
\label{subsec:inr}

\begin{figure*}[!t]
    \centering
    \includegraphics[width=1\linewidth]{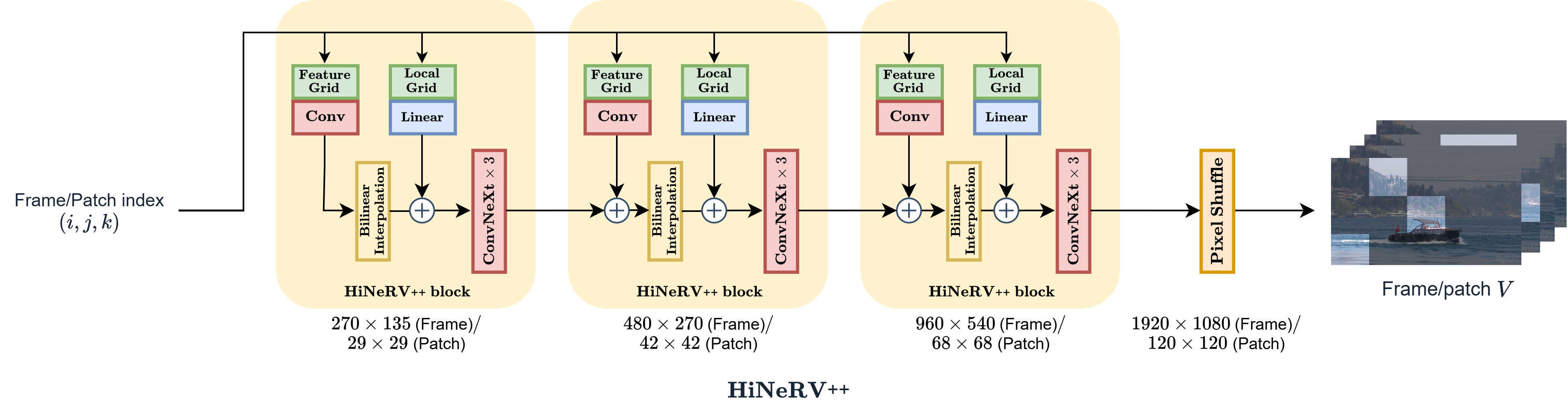}
    \caption{\textbf{The proposed HiNeRV++ architecture.} Following the original HiNeRV \cite{kwan2024hinerv}, the model can operate on either full frames or overlapping patches during both training and evaluation, while preserving strict train-test consistency.}
    \label{fig:hinerv++}
\end{figure*}

Our proposed HiNeRV++ architecture is primarily based on the original HiNeRV \cite{kwan2024hinerv}, but it introduces several major modifications, including the use of higher-resolution features, multi-reference inputs, and a pixel shuffle output layer, as detailed in the main paper. The network employs ConvNeXt blocks \cite{liu2022conv} as its core network layers. Additionally, it retains the use of bilinear interpolation and hierarchical encoding during the upsampling stages, consistent with the original design \cite{kwan2024hinerv}. Regarding the proposed multi-reference inputs, the network extracts the interpolated slice with bilinear interpolation, along with 6 additional raw reference slices, to formulate the input encodings from the feature grids.

Additionally, our implementation incorporates several minor structural refinements. Following \cite{gao2025pnvc}, we employ affine transformations instead of simple addition. Furthermore, we apply linear projection layers for the shortcut connections when the input and output channel dimensions of the ConvNeXt \cite{liu2022conv} blocks do not match, and we utilize LayerScale \cite{touvron2021going, liu2022conv} with an initial scale of $1 \times 10^{-4}$, which we found improves the training stability especially for the largest variants.

The complete schematic of the proposed HiNeRV++ is shown in \cref{fig:hinerv++}.

\subsection{Hyperparameter configurations}
\label{subsec:hyperparameters}

\begin{table}[t]
\scriptsize
\centering
\caption{\textbf{Detailed architectural configurations for NVRC++ across different scales (S1--S4).} The grid dimensions are shared across all scales. The temporal dimension of the feature grids scales dynamically with the input video sequence length $T$, yielding temporal lengths of $T$, $\lceil T/2 \rceil$, $\lceil T/4 \rceil$, $\lceil T/8 \rceil$, and $\lceil T/16 \rceil$. The specific grid dimensions shown here illustrate an example for a sequence with $T=600$ frames. For each stage, we report the output channels and local grid channels. Other configuration follows \cite{kwan2024hinerv, kwan2024nvrc}.}
\label{tab:hyperparameters}
\scriptsize
\setlength{\tabcolsep}{10pt}
\begin{tabular}{l l c c c c}
    \toprule
    \multirow{2}{*}{Stage} & \multirow{2}{*}{Grid Dimensions} & \multicolumn{4}{c}{Channels (Outputs / Local grids)} \\
    \cmidrule(lr){3-6}
    & & \textbf{S1} & \textbf{S2} & \textbf{S3} & \textbf{S4} \\
    \midrule
    \multirow{3}{*}{Stage 1} & $(\lceil T/4 \rceil, 45, 80, 4)$ & \multirow{3}{*}{$64$/$8$} & \multirow{3}{*}{$128$/$16$} & \multirow{3}{*}{$256$/$32$} & \multirow{3}{*}{$512$/$64$} \\
     & $(\lceil T/8 \rceil, 23, 40, 8)$ & & & & \\
     & $(\lceil T/16 \rceil, 12, 20, 16)$ & & & & \\
    \midrule
    Stage 2 & $(\lceil T/2 \rceil, 135, 240, 2)$ & $32$/$4$ & $64$/$8$ & $128$/$16$ & $256$/$32$ \\
    Stage 3 & $(T, 270, 480, 1)$ & $16$/$2$ & $32$/$4$ & $64$/$8$ & $128$/$16$ \\
    \midrule
    Output & \multicolumn{1}{c}{--} & $3$/-- & $3$/-- & $3$/-- & $3$/-- \\
    \bottomrule
\end{tabular}
\end{table}

\begin{table}[t]
\centering
\caption{\textbf{Feature grid masking configuration.} The activation ratio defines the proportion of feature channels keeping active during training. The low resolution grids (Stage 1) are always active. For the high resolution grids (Stages 2 and 3), the activation ratio is linearly annealed during specific phases of the training process, denoted here by the training progress (ratio of current epoch to total epochs). $T$ denote the input video sequence length  (see \cref{tab:hyperparameters}).}
\label{tab:masking_config}
\setlength{\tabcolsep}{10pt}
\scriptsize
\begin{tabular}{llcc}
    \toprule
    Stage & Grid Dimensions & Annealing Progress & Activation Ratio \\
    \midrule
    \multirow{3}{*}{Stage 1} & $(\lceil T/4 \rceil, 45, 80, 4)$ & \multirow{3}{*}{--} & \multirow{3}{*}{$1.0$} \\
     & $(\lceil T/8 \rceil, 23, 40, 8)$ & & \\
     & $(\lceil T/16 \rceil, 12, 20, 16)$ & & \\
    \midrule
    Stage 2 & $(\lceil T/2 \rceil, 135, 240, 2)$ & $0.2 \rightarrow 0.3$ & $0.01 \rightarrow 1.0$ \\
    Stage 3 & $(T, 270, 480, 1)$ & $0.3 \rightarrow 0.4$ & $0.01 \rightarrow 1.0$ \\
    \bottomrule
\end{tabular}
\end{table}

The different capacity/complexity variants of NVRC++ (S1--S4) are configured by scaling the size of the implicit neural representation (INR), which is achieved by adjusting the channel dimensions. The corresponding hyper-parameters are detailed in \cref{tab:hyperparameters}. We also provide the configuration of the proposed feature grid masking in \cref{tab:masking_config}.

\section{Experimental setup}
\label{sec:exp_setup}

\subsection{NVRC++}
To train NVRC++, we adopt the training configurations of the original NVRC \cite{kwan2024nvrc}, with a few key modifications. 

\vspace{5pt}\noindent\textbf{Encoding settings.} By default, we train our models for 720 and 1440 epochs on the UVG \cite{mercat2020uvg} and MCL-JCV \cite{wang2016mcl} datasets, respectively. We employ the Adam optimizer \cite{kingma2014adam} with a learning rate of $2 \times 10^{-3}$, a batch size of 1 (in frames), and a spatial patch size of $120 \times 120$. To estimate the bitrate, we randomly sample 20\% of the temporal slices from each level of the feature grids, but with a minimum of 32 slices selected.

Additionally, we apply the $\ell_2$ regularization weight of $1 \times 10^{-6}$, incorporating the annealing strategy \cite{kwan2024nvrc}.

\begin{figure}[!t]
    \centering
    \begin{subfigure}[t]{0.49\linewidth}
        \centering
        \vspace{0pt}
        \includegraphics[width=0.92\linewidth]{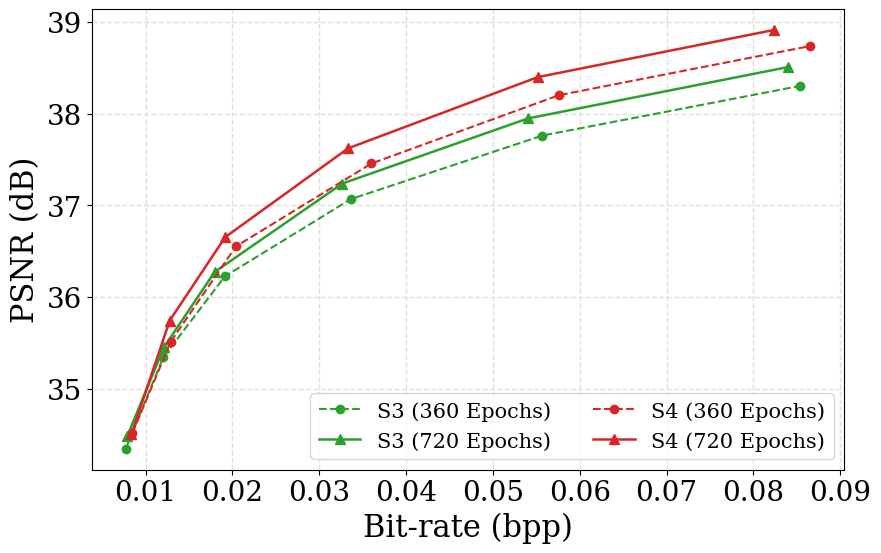}
    \end{subfigure}
    \begin{subfigure}[t]{0.49\linewidth}
        \centering
        \vspace{0pt}
        \includegraphics[width=0.9\linewidth]{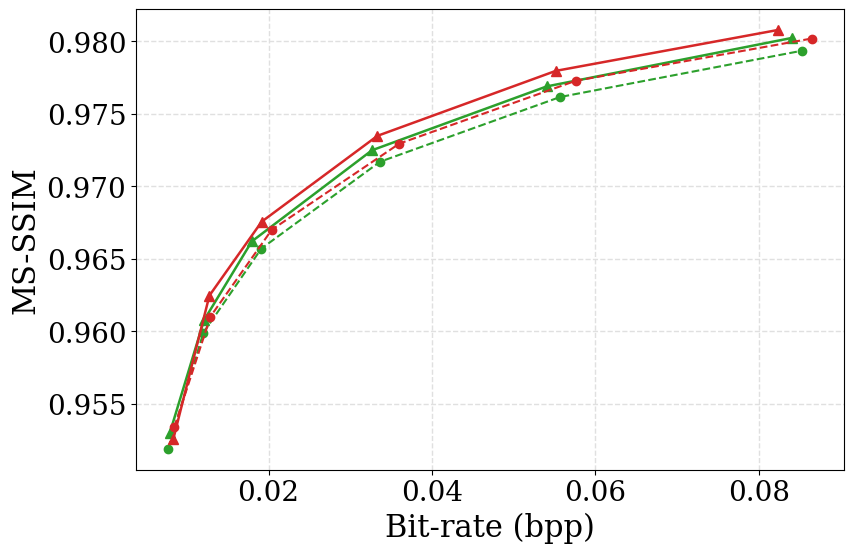} 
        \vspace{-15pt} 
    \end{subfigure}
    \caption{Comparison of NVRC++ (S3/S4) with different numbers of training epochs. The UVG dataset \cite{mercat2020uvg} is used. With 720 epochs of training, NVRC++ performance is further improved, while still maintaining a comparable coding cost with the original NVRC \cite{kwan2024nvrc}. Furthermore, even with only 360 epochs of training, NVRC++ already obtains highly competitive performance.}
    \label{fig:360e_vs_720e}
\end{figure}

Although this training process requires more epochs than the standard NVRC configuration, NVRC++ benefits from a significantly faster per-step optimization time compared to larger NVRC variants. This results in a comparable maximum encoding duration. \cref{fig:360e_vs_720e} presents a comparison with evaluating our method for fewer training epochs. This demonstrates that NVRC++ maintains highly competitive performance even under a constrained training budget.

\begin{figure}[!t]
    \centering
    \begin{subfigure}[t]{0.49\linewidth}
        \centering
        \vspace{0pt}
        \includegraphics[width=0.9\linewidth]{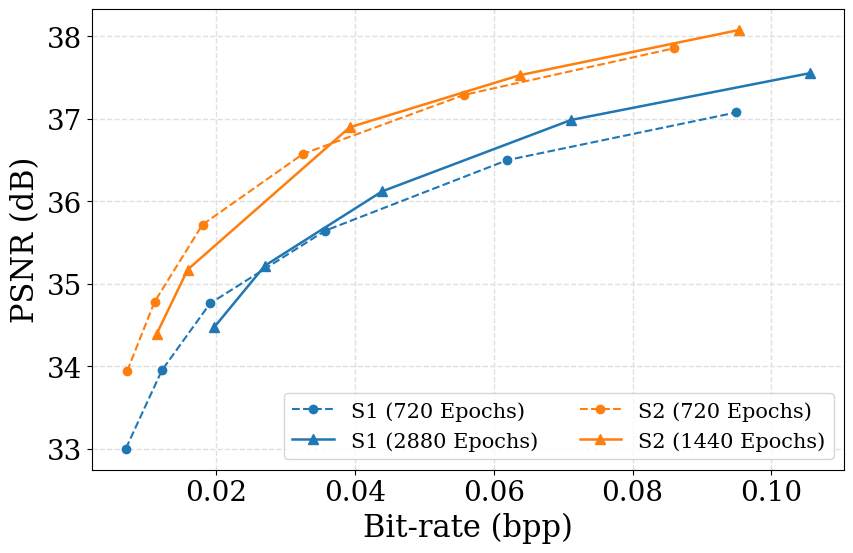}
    \end{subfigure}
    \begin{subfigure}[t]{0.49\linewidth}
        \centering
        \vspace{0pt}
        \includegraphics[width=0.9\linewidth]{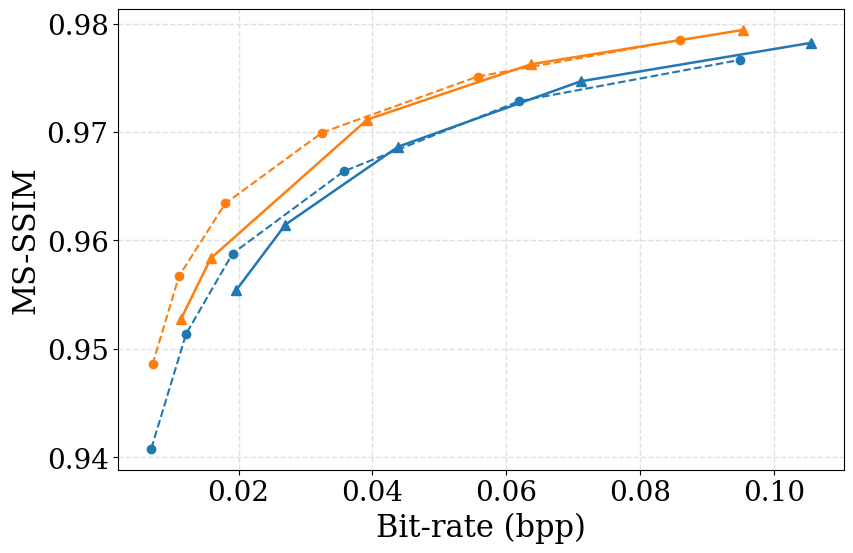} 
        \vspace{-15pt} 
    \end{subfigure}
    \caption{Comparison of NVRC++ (S1/S2) with different numbers of training epochs and video lengths. The UVG dataset \cite{mercat2020uvg} is used. With 2880/1440 epochs of training and splitting the video frames into groups of 75/150, NVRC++ performance is improved at higher bitrates. Note that even with much longer training, NVRC++ (S1/S2) still maintains a comparable encoding time to the original NVRC \cite{kwan2024nvrc} at large scales, but with a much faster decoding speed.}
    \label{fig:720e_vs_1440e_2880e}
\end{figure}

For the smaller model variants (S1 and S2) operating at high bitrates (the 4th to 6th rate points), we partition the longer videos \cite{mercat2020uvg} into shorter groups of 75 frames and 150 frames, respectively. These segments are then trained for 2880 epochs and 1440 epochs. We found that this temporal partitioning improves reconstruction quality at high bitrates, likely mitigating the limited capacity of these lightweight networks to fit long, complex video sequences. 

Although partitioning introduces some additional coding overhead, the S1 and S2 variants are highly compact and have a much faster encoding time per optimization step. Partitioning the videos also enables parallelization in the encoding process, allowing more powerful GPUs or multi-GPU setups to be leveraged. Furthermore, this is a one-off cost, where the S1 and S2 variants are extremely efficient during decoding. \cref{fig:720e_vs_1440e_2880e} provides an ablation study that compares the S1 and S2 models with and without this partitioning strategy.

\vspace{5pt}\noindent\textbf{Rate-Distortion Optimization.} Following previous works \cite{li2022nerv, kwan2024hinerv, kwan2024nvrc}, we employ a combination of $\ell_1$ and MS-SSIM losses as the training objective. To govern the rate-distortion trade-off, the multiplier $\lambda$ is set to $\{1.0, 2.0, 4.0, 8.0, 16.0, 32.0\}$ for the main experiments. For the ablation studies, we utilize a sparser set of $\lambda \in \{1.0, 3.2, 10.2, 32.8\}$ to efficiently cover a broad range of reconstruction qualities with reduced training cost. We decouple the optimization of the rate and distortion terms \cite{kwan2024nvrc}, updating the rate component once for every 8 optimization steps applied to the distortion term.

\vspace{5pt}\noindent\textbf{Quantization Approximation.} To optimize the network parameters under the rate-distortion objective, it is necessary to emulate the effects of hard quantization during training. Following the original NVRC \cite{kwan2024nvrc}, we employ soft-rounding with additive noise \cite{agustsson2020universally, kim2023c3} for this purpose. 

Specifically, the feature grid parameters are optimized using a soft-rounding temperature that anneals from $0.5$ down to $0.1$. For all remaining parameters, including the network weights, the entropy model parameters, and the quantization parameters, we utilize the temperature that is annealed from $0.5$ to $0.3$. We found that having different settings for the grid and weight parameters is helpful for the optimization.

\vspace{5pt}\noindent\textbf{$\ell_2$ regularization.} While the original NVRC framework applies $\ell_2$ regularization to the network parameters, we specifically omit $\ell_2$ regularization for the feature grids. We empirically found that penalizing the grid parameters in this manner restricts the model's representational capacity, which ultimately leads to underfitting at high bitrates.

\vspace{5pt}\noindent\textbf{Removal of the fine-tuning stage.} To reduce computational overhead and maintain pipeline simplicity, we omit the second-stage fine-tuning \cite{kwan2024nvrc}. Our empirical evaluations demonstrate that this additional stage yields only marginal performance improvements.

\vspace{5pt}\noindent\textbf{Encoding and decoding time.} We report the encoding and decoding time of NVRC++ using an NVIDIA RTX 4090 GPU, with the default configuration (720 training epochs). We utilize \texttt{FP16} during training, and \texttt{FP16} during evaluation for frame reconstruction, but use \texttt{FP32} for entropy encoding and decoding.

\subsection{Baseline Methods}
For conventional video codecs (x265 \cite{x265}, HM \cite{hm}, and VTM \cite{vtm}), we generate the results using their publicly available implementations. 

For the neural video compression baselines, we source the quantitative results directly from the original papers or their official repositories whenever possible (i.e., C3 \cite{kim2023c3} for the UVG dataset, and HiNeRV \cite{kwan2024hinerv}). Where pre-computed results are unavailable, we evaluate the models ourselves using the official codebases and pre-trained checkpoints provided by the authors (i.e., DCVC-FM \cite{li2024neural}, DCVC-RT \cite{jia2025towards}, C3 \cite{kim2023c3} for the MCL-JCV dataset, and NVRC \cite{kwan2024nvrc}).

Finally, we report the encoding and decoding latencies of the baseline neural compression models, where the results are obtained on the same platform (an NVIDIA RTX 4090 GPU with FP16).

\section{Extended quantitative results}
\label{sec:quant_results}

\subsection{Detailed rate-distortion results}
\label{subsec:full_rd_results}

We provide the detailed rate-distortion plots for better evaluation, as shown in \cref{fig:rdcurve_uvg} and \cref{fig:rdcurve_mcl}.

\subsection{Performance analysis of individual scales}
\label{subsec:individual_scales}

We provide additional discussion regarding the performance highlight of each variant of NVRC++.

\vspace{5pt}\noindent\textbf{S1.}
Our lightest variant, NVRC++ (S1), achieves ultra-low complexity (7.3 kMACs/pixel), making it one of the lowest-complexity models in the literature, while enabling up to 365.0 FPS for frame reconstruction at 1920 $\times$ 1080 resolution.

Compared to the C3 \cite{kim2023c3}, a SOTA lightweight overfitted codec with comparable complexity, NVRC++ (S1) demonstrates significantly better compression performance. Note that unlike C3, which is optimized for MSE loss, NVRC++ (S1) is jointly optimized for $\ell_1$ and MS-SSIM loss. On the UVG dataset \cite{mercat2020uvg}, it achieved 10.95\% / 58.17\% BD-rate gains (PSNR/MS-SSIM) over C3 (as the anchor). Furthermore, C3 requires a costly parameter searching process during encoding and utilizes a slow autoregressive model for decoding, whereas our model maintains a single, highly efficient configuration for both high-quality encoding and very fast actual decoding speed.

Furthermore, NVRC++ (S1) yields comparable MS-SSIM performance to HiNeRV \cite{kwan2024hinerv} (which is optimized for the same objective), achieving 0.28\% BD-rate (MS-SSIM) gain when using HiNeRV as the anchor. Compared to DCVC-RT \cite{jia2025towards} (optimized for MSE loss), it achieves a 10.53\% BD-rate (MS-SSIM) gain. Most notably, NVRC++ (S1) reaches these milestones at a drastically lower computational complexity of 7.3 kMACs/pixel, compared to 87.3--346.3 kMACs/pixel by HiNeRV and 166.8 kMACs/pixel by DCVC-RT.

\vspace{5pt}\noindent\textbf{S2.} Compared to S1, our S2 variant introduces a modest increase in computational complexity (approximately 3.4 times higher), while delivering substantially stronger coding performance. Specifically, it achieves BD-rate (PSNR/MS-SSIM) gains of 46.78\% / 33.11\% over S1 on the UVG dataset \cite{mercat2020uvg}.

\begin{figure}[!t]
    \centering
    \begin{subfigure}[t]{1.0\linewidth}
        \centering
        \includegraphics[width=0.8\linewidth]{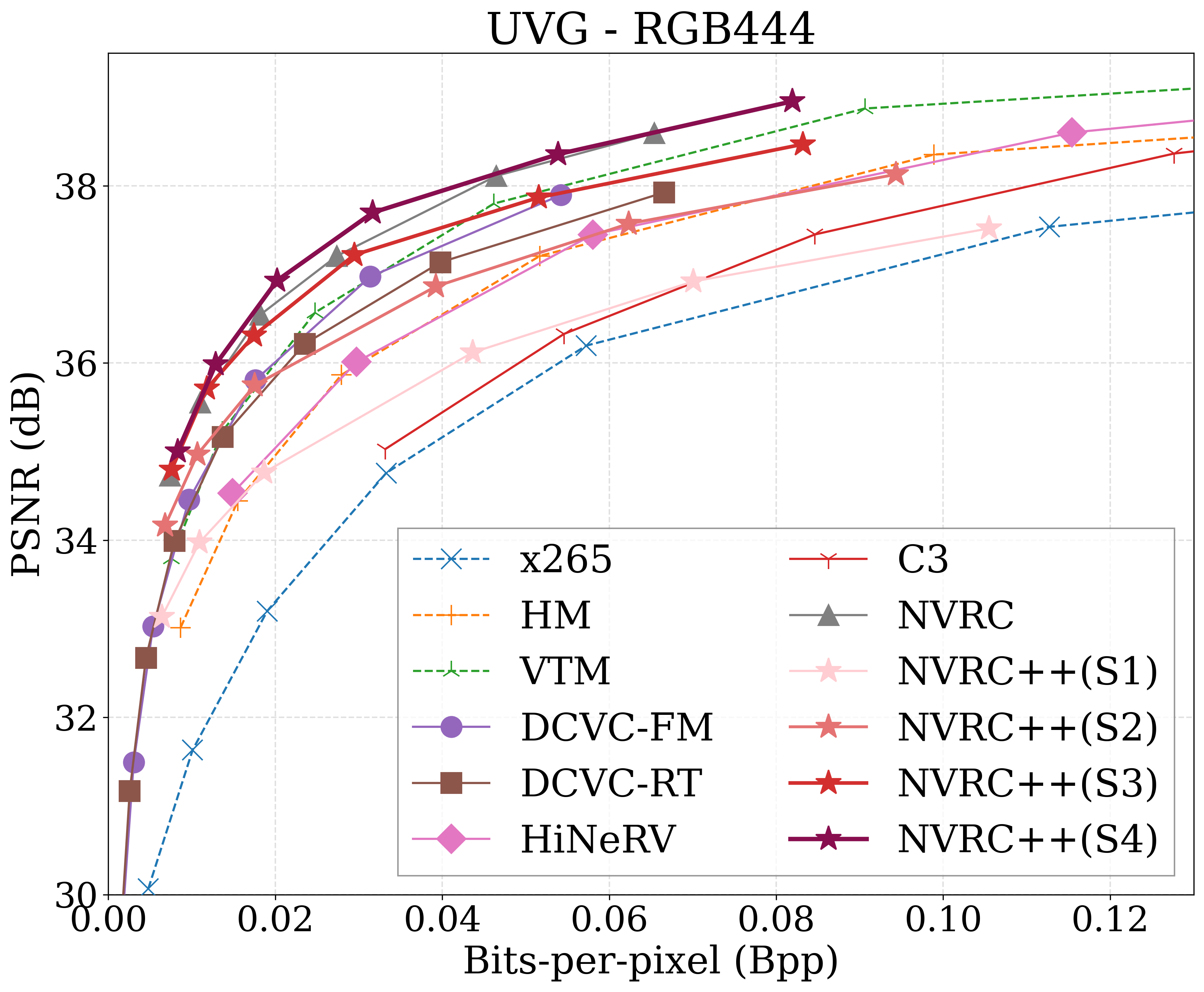}
    \end{subfigure}
    \begin{subfigure}[t]{1.0\linewidth}
        \centering
        \includegraphics[width=0.8\linewidth]{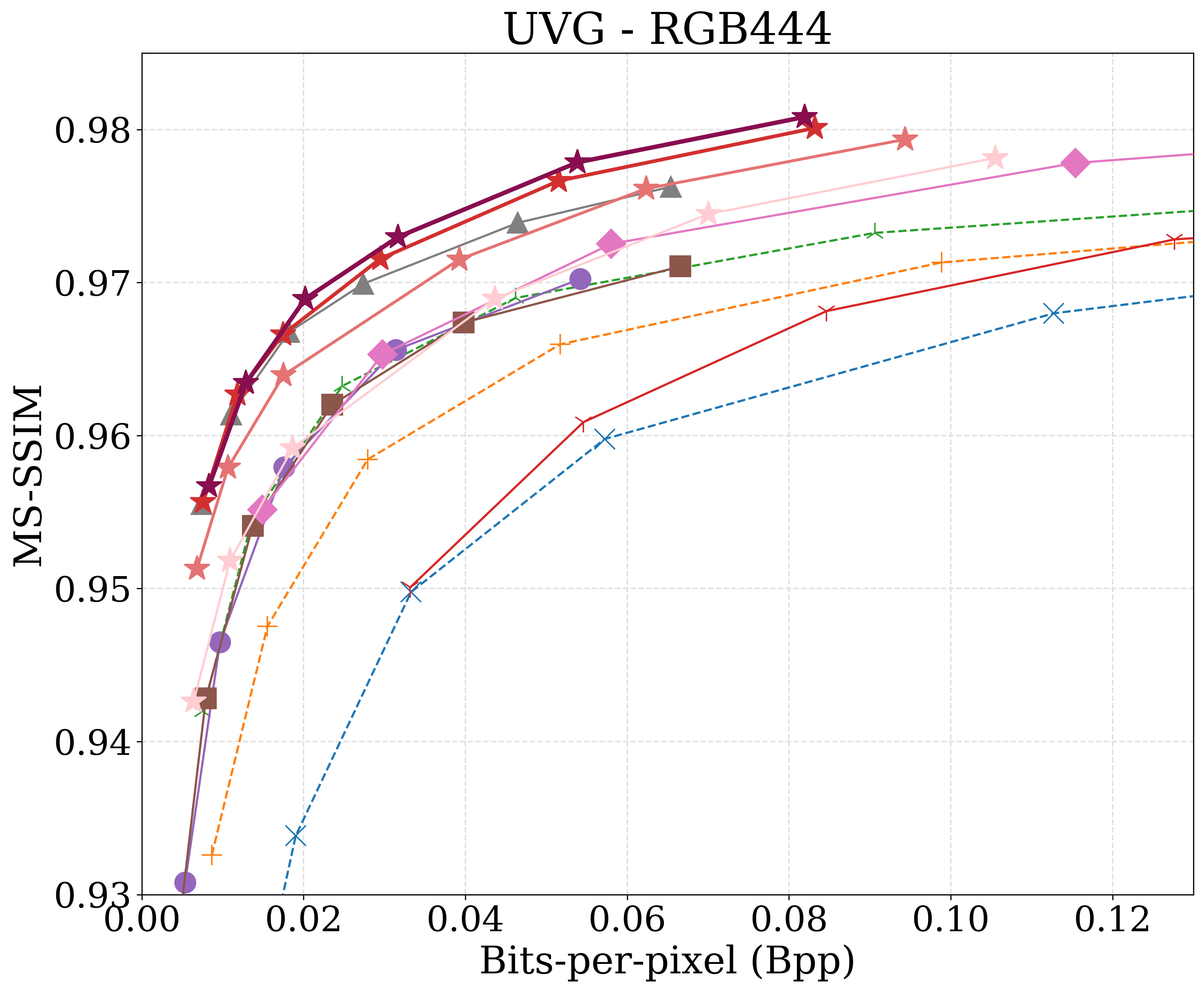} 
        \vspace{-15pt} 
    \end{subfigure}
    \caption{Detailed rate-distortion results with the UVG dataset \cite{mercat2020uvg}.}
    \label{fig:rdcurve_uvg}
\end{figure}

\begin{figure}[!t]
    \centering
    \begin{subfigure}[t]{1.0\linewidth}
        \centering
        \includegraphics[width=0.8\linewidth]{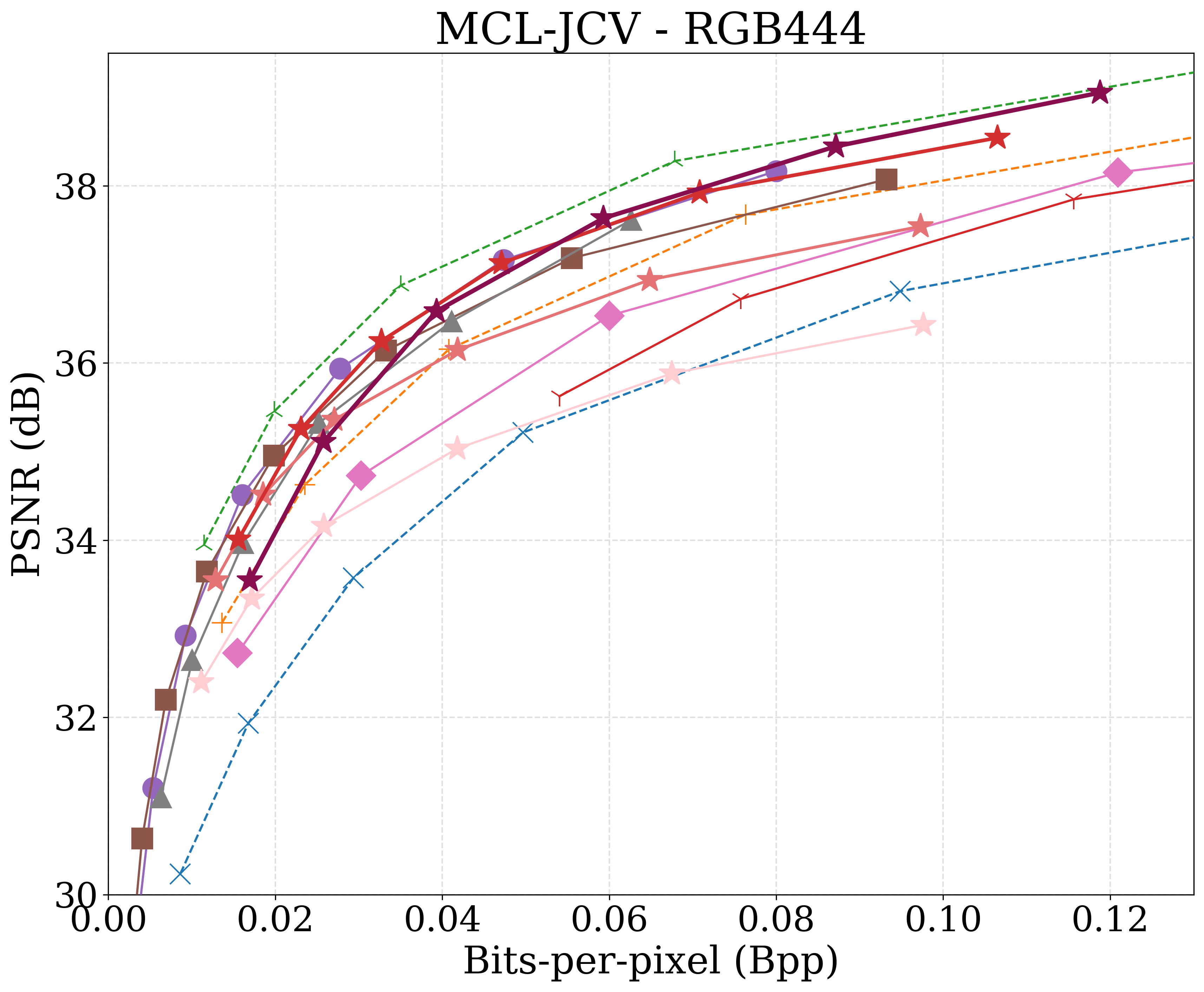}
    \end{subfigure}
    \begin{subfigure}[t]{1.0\linewidth}
        \centering
        \includegraphics[width=0.8\linewidth]{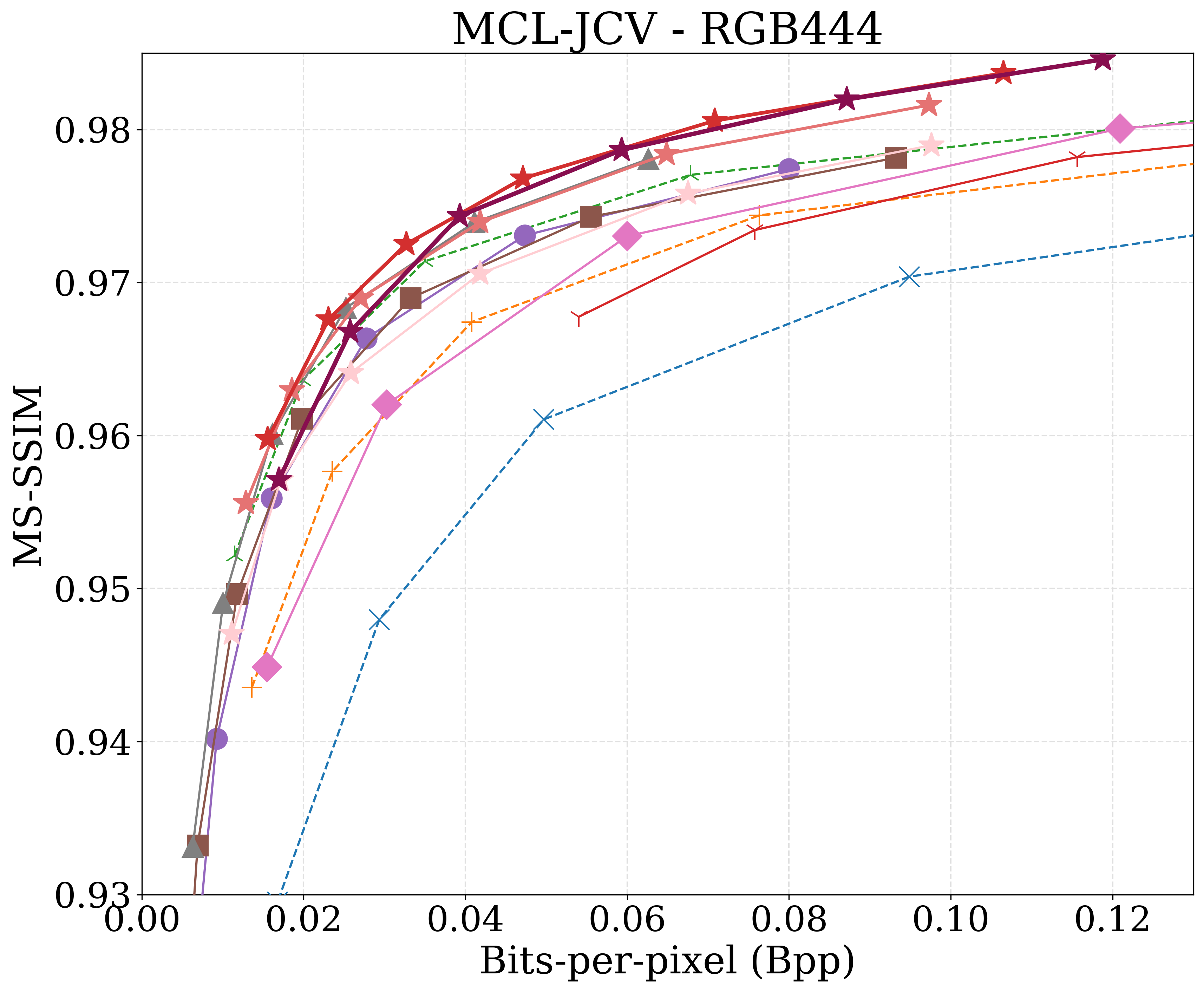} 
        \vspace{-15pt} 
    \end{subfigure}
    \caption{Detailed rate-distortion results with the MCL-JCV dataset \cite{wang2016mcl}.}
    \label{fig:rdcurve_mcl}
\end{figure}

As illustrated in \cref{fig:rdcurve_uvg}, S2 outperforms HM (Random Access) \cite{hm}, HiNeRV \cite{kwan2024hinerv}, and even SOTA neural codecs such as DCVC-FM \cite{li2024neural} and DCVC-RT \cite{jia2025towards} at low bitrates on the UVG dataset in terms of PSNR. Furthermore, it matches the performance of VTM (Random Access) \cite{vtm} at certain rate points on the UVG dataset (\cref{fig:rdcurve_uvg}), as well as the original NVRC \cite{kwan2024nvrc} on the MCL-JCV dataset \cite{wang2016mcl} (\cref{fig:rdcurve_mcl}).

Compared to HiNeRV, NVRC++ (S2) achieves BD-rate (PSNR/MS-SSIM) improvements of 22.59\% / 32.76\% on the UVG dataset and 23.88\% / 38.98\% on the MCL-JCV dataset. Remarkably, it reaches these performance levels while requiring only 7\% of the computational complexity (measured in kMACs/pixel) of HiNeRV. This highlights the exceptionally high efficiency of our proposed architecture. Notably, S2 reconstructs $1920 \times 1080$ frames at 162.9 FPS, ranking it among the fastest models in the literature.

\vspace{5pt}\noindent\textbf{S3.} NVRC++(S3) has achieved SOTA-level compression performance while maintaining a lightweight complexity. Compared to S2, it achieves BD-rate savings (PSNR/MS-SSIM) of 33.34\% / 22.91\% and 19.52\% / 9.00\% on the UVG \cite{mercat2020uvg} and MCL-JCV \cite{wang2016mcl} datasets, respectively.

While jointly optimized for $\ell_1$ and MS-SSIM, it consistently outperforms strong baselines including VTM (Random Access) \cite{vtm}, DCVC-FM \cite{li2024neural}, and DCVC-RT \cite{jia2025towards} in terms of PSNR on the UVG dataset. Furthermore, it significantly outperforms NVRC \cite{kwan2024nvrc} at high bitrates when measured in MS-SSIM, as shown in \cref{fig:rdcurve_uvg}.

Quantitatively, NVRC++(S3) attains BD-rate (PSNR/MS-SSIM) reduction of 30.51\% / 54.93\% and -3.97\% / 29.61\% against the SOTA DCVC-RT on the UVG and MCL-JCV datasets, respectively. Notably, it operates at a complexity of only 92.5 kMACs/pixel. This is significantly lower than NVRC (173.4--930.6 kMACs/pixel), DCVC-FM (369.6 / 865.5 kMACs/pixel for I/P-frames), and DCVC-RT (364.9 / 166.8 kMACs/pixel for I/P-frames), and achieving a real-time decoding speed of 73.4 FPS.

Furthermore, S3 outperformed our largest variant, S4, on shorter video sequences (MCL-JCV) at low bitrates. This suggests that the extra network parameter count and complexity may not be helpful for capturing the redundancy in short video sequences.

\vspace{5pt}\noindent\textbf{S4.} Our largest variant, S4, further pushes the compression limit on long video sequences (UVG \cite{mercat2020uvg}), achieving additional BD-rate (PSNR/MS-SSIM) savings of 17.10\% / 4.60\% over S3. This indicates the high efficiency of S4 for capturing the spatial-temporal redundancy in long video sequences.

Compared to the SOTA INR-based method NVRC \cite{kwan2024nvrc}, our S4 achieves comparable PSNR performance and superior MS-SSIM performance, yielding BD-rate savings of 8.00\% / 10.73\%. Furthermore, NVRC++ (S4) operates at a fixed computational complexity of 357.7 kMACs/pixel, avoiding the variable, bitrate-dependent overhead (173.4--930.6 kMACs/pixel) seen in NVRC. Even as our largest model variant, S4 maintains a real-time decoding speed of 34 FPS.

Compared to other SOTA codecs, S4 obtains BD-rate savings(PSNR/MS-SSIM) of 27.71\% / 58.17\% over VTM (Random Access) \cite{vtm}, 33.75\% / 54.06\% over DCVC-FM \cite{li2024neural}, and 40.30\% / 55.52\% over DCVC-RT \cite{jia2025towards} on the UVG dataset. These results validate our framework's superior performance and the flexibility to scale down to lower complexities (S1--S3) while maintaining real-time throughput.

\subsection{Encoding and decoding latency}
\label{subsec:latency}
While achieving state-of-the-art compression performance alongside real-time decoding speeds, our architecture requires a lengthy encoding time due to the overfitting paradigm. This remains an inherent limitation of all overfitted, INR-based codecs. However, we note that encoding is a one-off cost. Because these models do not require generalization, INR-based codecs can ultimately provide a vastly superior trade-off between decoding performance and computational complexity.

As detailed in \cref{sec:exp_setup}, encoding and frame reconstruction times are measured using \texttt{FP16} precision, while entropy coding is measured using \texttt{FP32}. Further computational acceleration could be achieved by adopting \texttt{INT8}/\texttt{FP8} or lower-precision formats for both processes. While entropy decoding time in our current implementation could present a bottleneck for real-world deployment, especially for the lightest variant (S1/S2), our entropy decoding complexity remains exceptionally low (approximately 0.3 kMACs/pixel). Therefore, optimizing the engineering implementation or overlapping the entropy decoding step with frame reconstruction \cite{jia2025towards} could effectively mitigate this overhead.

Furthermore, the recent neural codec DCVC-RT \cite{jia2025towards} has demonstrated real-time encoding and decoding performance while achieving SOTA compression. We note that this very fast decoding speed is partially due to the use of customized CUDA kernels and a parallel coding pipeline, whereas our implementation is based entirely on standard PyTorch and does not utilize parallel coding. We measured the performance of DCVC-RT with and without its customized kernels on the same hardware platform as ours. While DCVC-RT achieves a 115.76 FPS decoding speed with the kernels, it drops to only 50.13 FPS when plain PyTorch code is used. This is slower than the frame reconstruction speed of our S1--S3 models (73.4 to 365.0 FPS). Implementing customized kernels could further improve our model's decoding speed.

\subsection{Memory consumption analysis}
\label{subsec:memory_usage}

\begin{table}[htbp]
\centering
\scriptsize
\caption{Peak GPU memory usage (MB) for NVRC++ variants. Results are measured at $1920 \times 1080$ resolution for a sequence of 600 frames with \texttt{FP16} precision.}
\label{tab:memory_usage}
\begin{tabular}{@{}lcc@{}}
\toprule
\textbf{Model Variant} & \textbf{Encoding (MB)} & \textbf{Decoding (MB)} \\ \midrule
NVRC++ (S1)              & 8645.79                & 6318.71                \\
NVRC++ (S2)              & 8659.04                & 6328.99                \\
NVRC++ (S3)              & 13309.38               & 6359.92                \\
NVRC++ (S4)              & 21476.08                & 6466.05                    \\ \bottomrule
\end{tabular}
\end{table}

In \cref{tab:memory_usage}, we provide the peak memory usage for both encoding and decoding. All model variants fit comfortably within a standard consumer-grade 24 GB GPU. It should be noted that the training memory footprint is based on our default configuration; however, since our model utilizes patch-level optimization \cite{kwan2024hinerv} and our proposed feature grid slice sampling, the memory requirements can be further reduced by using a smaller training batch size. During inference, all variants exhibit nearly identical memory costs. This is because the memory footprint is dominated by the entropy models, which are consistent across all variants. Furthermore, we currently utilize a large batch size for entropy coding to support hierarchical B-frame style coding of the feature grids; this memory overhead can be significantly mitigated by reducing the degree of parallelism during the entropy coding stage.

\subsection{Bitrate allocation across model parameters}
\label{subsec:rate_distribution}

\begin{figure}[!t]
    \centering
    \includegraphics[width=0.99\linewidth]{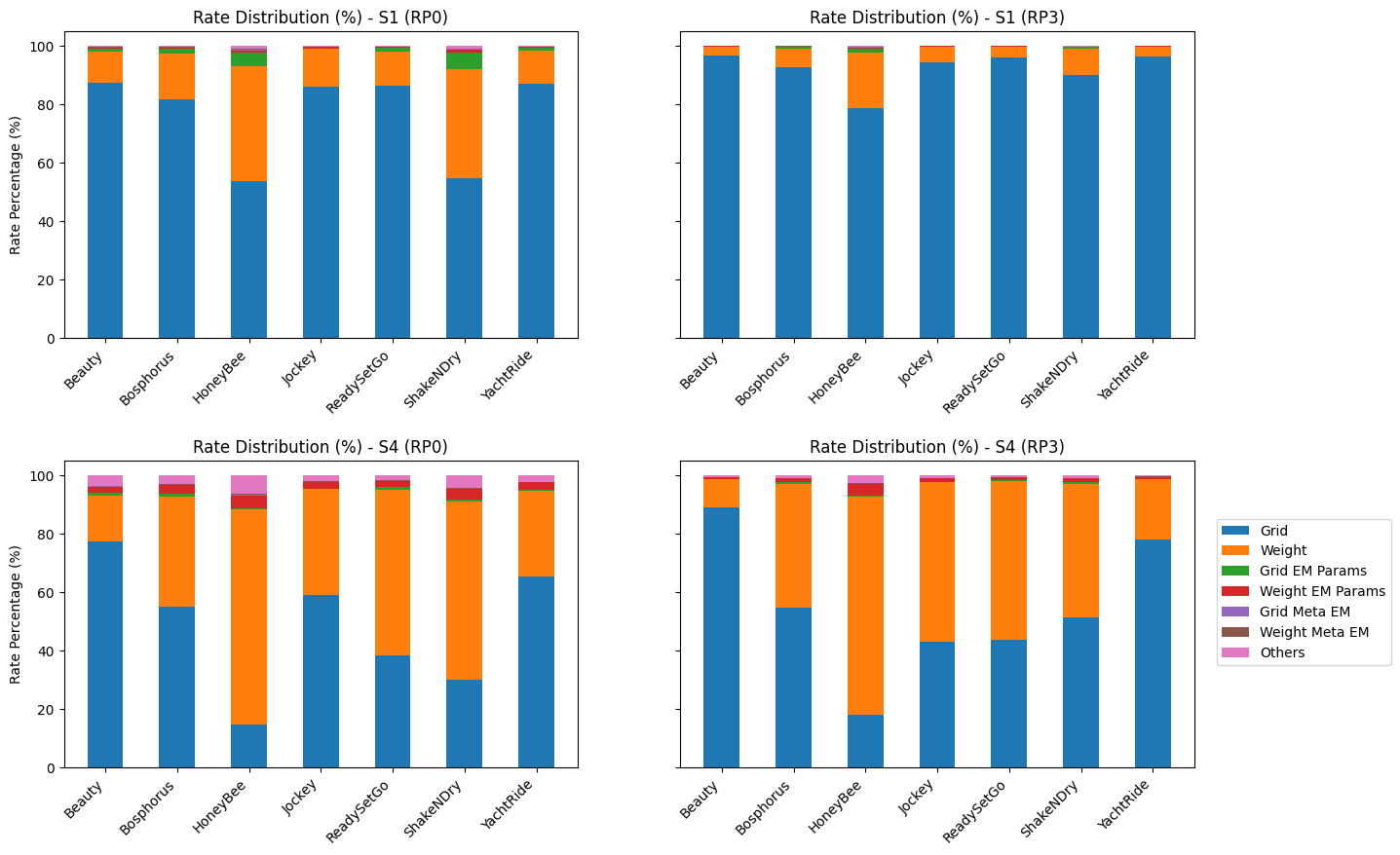}
    \caption{Per-sequence bitrate distribution on the UVG dataset \cite{mercat2020uvg} for different variants (S1 and S4) and rate points (the first and fourth). This illustrates that bitrates are dynamically allocated in an end-to-end manner, varying significantly across different sequences and rate points.}
    \label{fig:rate_dist}
   \vspace{-15pt}
\end{figure}

We present the bitrate distribution for the S1 and S4 models at two distinct rate points in \cref{fig:rate_dist}. The results demonstrate that bitrate allocation is dynamic and achieved by the end-to-end optimization process \cite{kwan2024nvrc}. Specifically, highly dynamic sequences (e.g., \textit{Beauty}) receive a higher bit allocation for the feature grids, whereas static sequences (e.g., \textit{HoneyBee}) allocate more bits to the network layer parameters. Furthermore, the smaller model (S1) allocates a larger portion of the bit budget to the grids; this is expected, as high-resolution grids improve performance by storing high-quality features. Finally, both variants allocate a higher proportion of bits to the feature grids at higher rate points, indicating their increased importance for high-quality reconstruction. This validates our motivation: high-resolution grids provide superior scalability and are essential for achieving the rate-distortion performance required at higher bitrates.

\subsection{Temporal distortion variations}
\label{subsec:framewise_distortion}

\begin{figure}[!t]
    \centering
    \begin{subfigure}[t]{0.48\linewidth}
        \centering
        \includegraphics[width=0.99\linewidth]{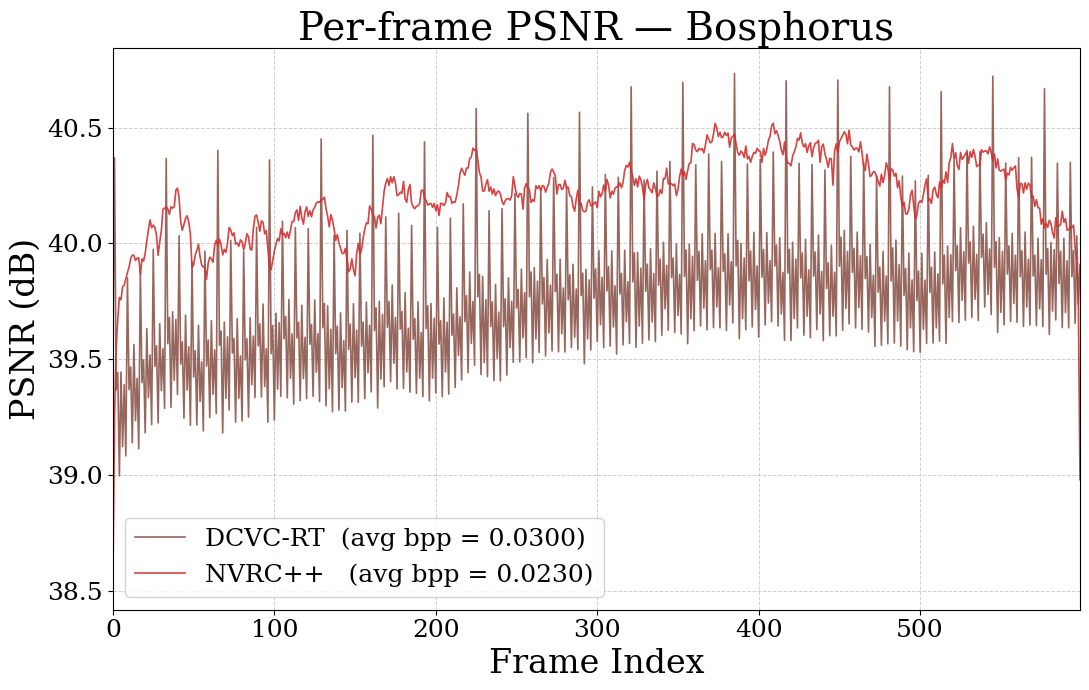}
    \end{subfigure}
    \begin{subfigure}[t]{0.48\linewidth}
        \centering
        \includegraphics[width=0.99\linewidth]{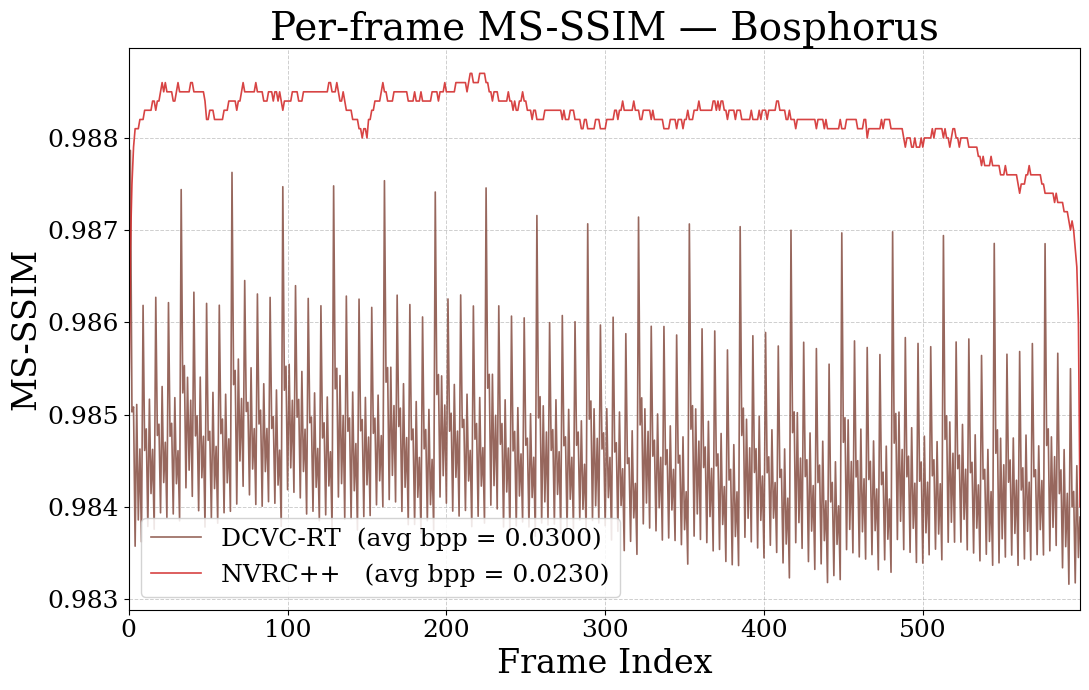}
    \end{subfigure}
    \begin{subfigure}[t]{0.48\linewidth}
        \centering
        \includegraphics[width=0.99\linewidth]{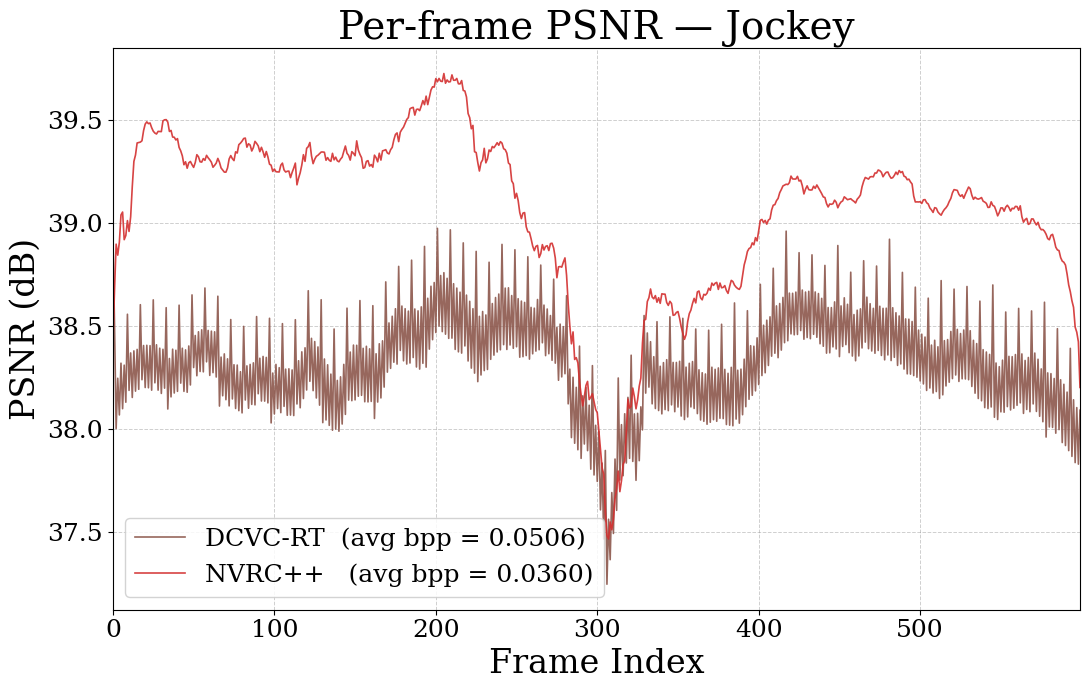} 
        \vspace{-15pt} 
    \end{subfigure}
    \begin{subfigure}[t]{0.48\linewidth}
        \centering
        \includegraphics[width=0.99\linewidth]{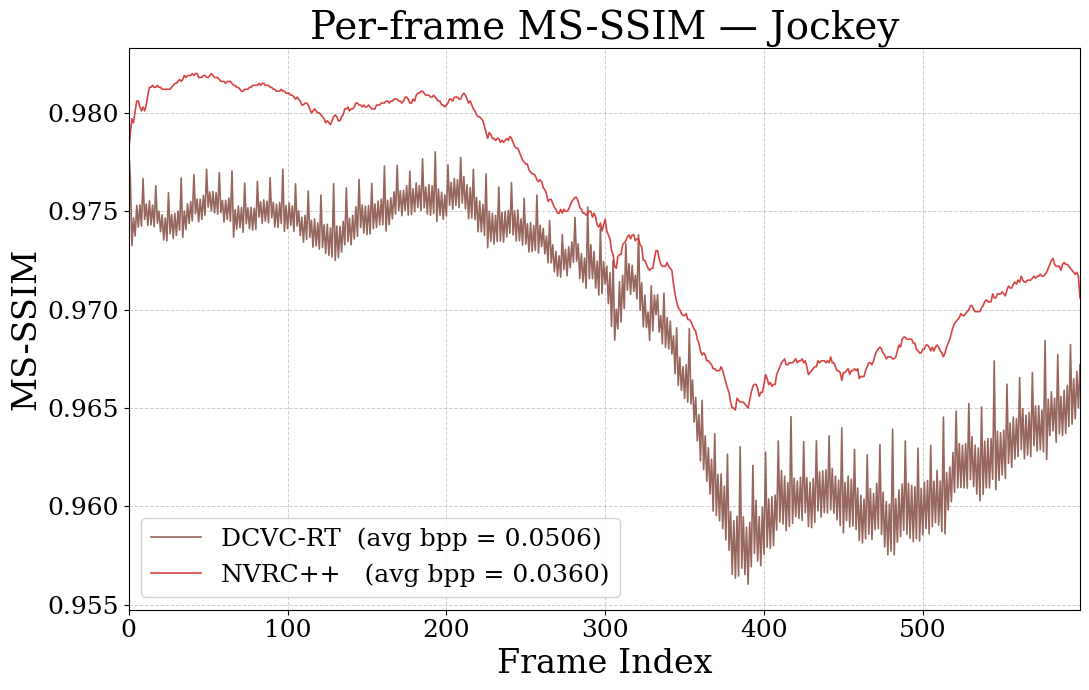}
        \vspace{-15pt} 
    \end{subfigure}
    \caption{Per-frame distortion results on the UVG dataset (\textit{Bosphorus} and \textit{Jockey} sequences) \cite{mercat2020uvg}. Compared to DCVC-RT \cite{jia2025towards}, the proposed NVRC++ (S4) exhibits a significantly smoother distortion curve across frames. This stability stems from the joint optimization of all frames during the encoding process, which effectively mitigates the quality fluctuations often seen in frame-by-frame predictive coding.}
    \label{fig:temp_distortion}
   \vspace{-15pt} 
\end{figure}

In \cref{fig:temp_distortion}, we present the distortion performance over time for both DCVC-RT \cite{jia2025towards} and our proposed NVRC++. Compared to DCVC-RT, NVRC++ exhibits a significantly smoother temporal distortion curve. This stability arises because NVRC++ jointly optimizes all frames to achieve strong performance for coding long videos. In contrast, DCVC-RT relies on manually designed hierarchical quality mechanisms \cite{li2024neural, jia2025towards}, which causes substantial quality fluctuations between adjacent frames.

\section{Additional results on JVET-CTC class B dataset}
\label{sec:jvet-ctc_b}
We provide additional results on the JVET-CTC Class B dataset \cite{boyce2018jvet}, using the experimental setting in the main paper. We compare the proposed NVRC++ with several baselines, including x265 (\textit{veryslow}) \cite{x265}, HM-18.0 (\textit{randomaccess}) \cite{hm}, VTM-20.0 (\textit{randomaccess}) \cite{vtm}, and NVRC \cite{kwan2024nvrc}. The results are provided in \cref{tab:bdrate_classb}. Consistent with the main paper's results, the proposed NVRC++ obtains state-of-the-art performance on the JVET-CTC Class B dataset, where the largest variant (S4) outperforms both VTM (\textit{RA}) and NVRC.

\begin{table}[t]
    \centering
    \caption{BD-rate results of the proposed NVRC++ model (configured at four scales S1--S4) against conventional and neural video codecs on the JVET-CTC Class B dataset. BD-rate is measured against x265 (\textit{veryslow}).}
    \resizebox{0.4\linewidth}{!}{
    \begin{tabular}{lcc}
        \toprule
        \multicolumn{1}{r}{BD-rate (\%)} & \multicolumn{2}{c}{JVET-CTC Class B} \\
        \cmidrule(lr){2-3}
        Method & PSNR & MS-SSIM \\
        \midrule
        x265 (\textit{veryslow}) \cite{x265} & 0.00 & 0.00 \\
        HM (\textit{RA}) \cite{hm}           & -45.42 & -49.20 \\
        VTM (\textit{RA}) \cite{vtm}         & -63.53 & -69.23 \\
        \midrule
        NVRC \cite{kwan2024nvrc}             & -68.70 & -84.69 \\
        \midrule
        \textbf{NVRC++ (S1)}                 & 38.72 & -41.45 \\
        \textbf{NVRC++ (S2)}                 & -23.42 & -66.39 \\
        \textbf{NVRC++ (S3)}                 & -60.06 & -81.34 \\
        \textbf{NVRC++ (S4)}                 & -70.94 & -85.93 \\
        \bottomrule
    \end{tabular}}
    \label{tab:bdrate_classb}
\end{table}

\section{Additional results on YUV420 color space}
\label{sec:YUV}
We provide additional results in the YUV420 color space on the UVG \cite{mercat2020uvg}, MCL-JCV \cite{wang2016mcl}, and JVET-CTC Class B \cite{boyce2018jvet} datasets. All experiments are conducted on the original YUV source, using PSNR-YUV (6:1:1) and MS-SSIM (Y) as the metrics. We compare the proposed NVRC++ with several baselines, including x265 (\textit{veryslow}) \cite{x265}, HM-18.0 (\textit{randomaccess}) \cite{hm}, VTM-20.0 (\textit{randomaccess}) \cite{vtm}, and NVRC \cite{kwan2024nvrc}. The results are provided in \cref{tab:bdrate_yuv420}. We use the same loss function as the original NVRC \cite{kwan2024nvrc}. The results show that NVRC++ performs well across different settings, with S3/S4 still reaching state-of-the-art performance.

\begin{table}[t]
    \centering
    \caption{BD-rate results of the proposed NVRC++ model (configured at four scales S1-S4) against conventional and neural video codecs on the UVG, MCL-JCV, and JVET-CTC Class B datasets, evaluated in the YUV420 color space. BD-rate is measured against x265 (\textit{veryslow}), using PSNR-YUV (6:1:1) and MS-SSIM (Y) as quality metrics.}
    \resizebox{1\linewidth}{!}{
    \begin{tabular}{lcccccc}
        \toprule
        \multicolumn{1}{r}{BD-rate (\%)} & \multicolumn{2}{c}{UVG} & \multicolumn{2}{c}{MCL-JCV} & \multicolumn{2}{c}{JVET-CTC Class B} \\
        \cmidrule(lr){2-3} \cmidrule(lr){4-5} \cmidrule(lr){6-7}
        Method & PSNR (YUV) & MS-SSIM (Y) & PSNR (YUV) & MS-SSIM (Y) & PSNR (YUV) & MS-SSIM (Y) \\
        \midrule
        x265 (\textit{veryslow}) \cite{x265} & 0.00 & 0.00 & 0.00 & 0.00 & 0.00 & 0.00 \\
        HM (\textit{RA}) \cite{hm}           & -38.59 & -25.85 & -36.67 & -26.98 & -41.10 & -33.53 \\
        VTM (\textit{RA}) \cite{vtm}         & -58.00 & -48.35 & -56.51 & -49.99 & -59.56 & -58.61 \\
        \midrule
        NVRC \cite{kwan2024nvrc}             & -66.89 & -59.38 & -49.02 & -43.00 & -66.33 & -66.02 \\
        \midrule
        \textbf{NVRC++ (S1)}                 & -28.92 & -30.51 & -14.72 & -21.43 & 48.33 & 10.33 \\
        \textbf{NVRC++ (S2)}                 & -54.22 & -50.87 & -41.92 & -36.59 & -16.51 & -28.67 \\
        \textbf{NVRC++ (S3)}                 & -64.79 & -60.83 & -50.14 & -40.88 & -53.87 & -58.28 \\
        \textbf{NVRC++ (S4)}                 & -67.02 & -59.66 & -43.72 & -18.69 & -66.44 & -69.36 \\
        \bottomrule
    \end{tabular}}
    \label{tab:bdrate_yuv420}
\end{table}

\section{Further results on scalable coding}
\label{sec:scalable_coding}

Although NVRC++ is not explicitly optimized for scalable video coding, it inherently supports this capability through our masking mechanism. Specifically, by masking the high-resolution feature grids during the early stages of training, the model is forced to reconstruct video frames relying solely on the lower-resolution grids. Even as this masking is annealed over the course of training, the network retains the ability to produce high-quality reconstructions using only a partial subset of the latent representations. 

\begin{figure}[t]
    \centering
    \begin{subfigure}[b]{1.0\linewidth}
        \centering
        \includegraphics[width=\linewidth]{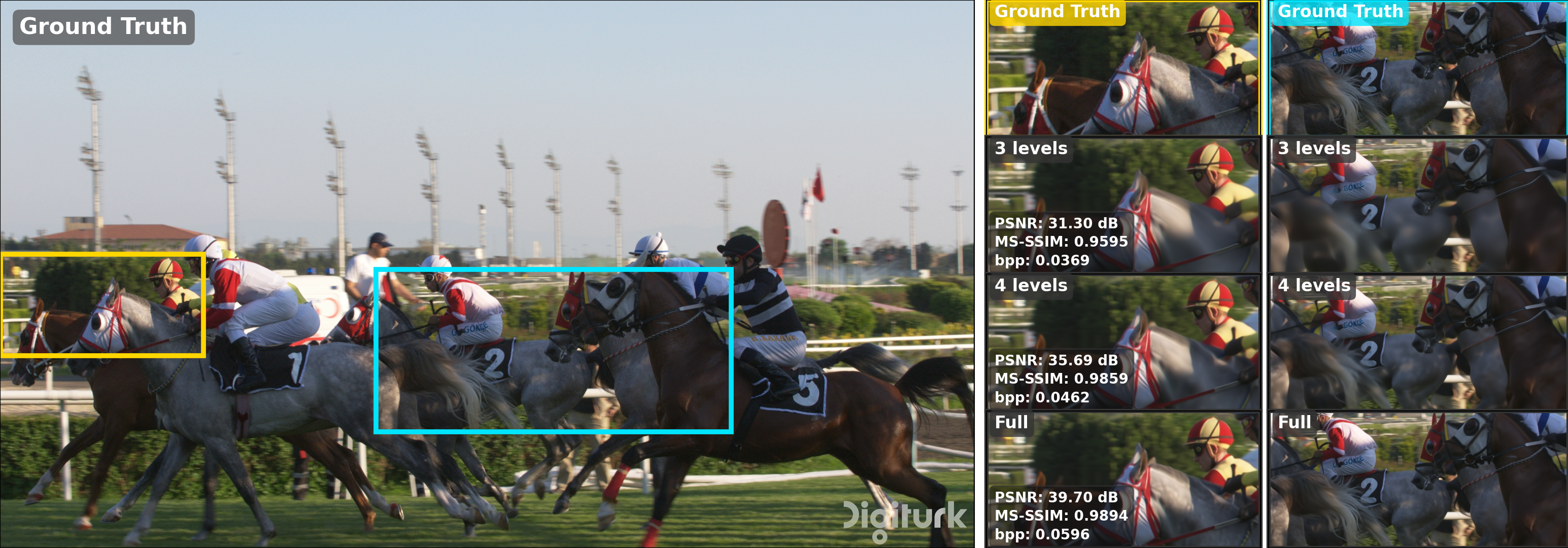}
        \caption{UVG - ReadySetGo}
        \label{fig:scalable_uvg}
    \end{subfigure}
    \hfill %
    \begin{subfigure}[b]{1.0\linewidth}
        \centering
        \includegraphics[width=\linewidth]{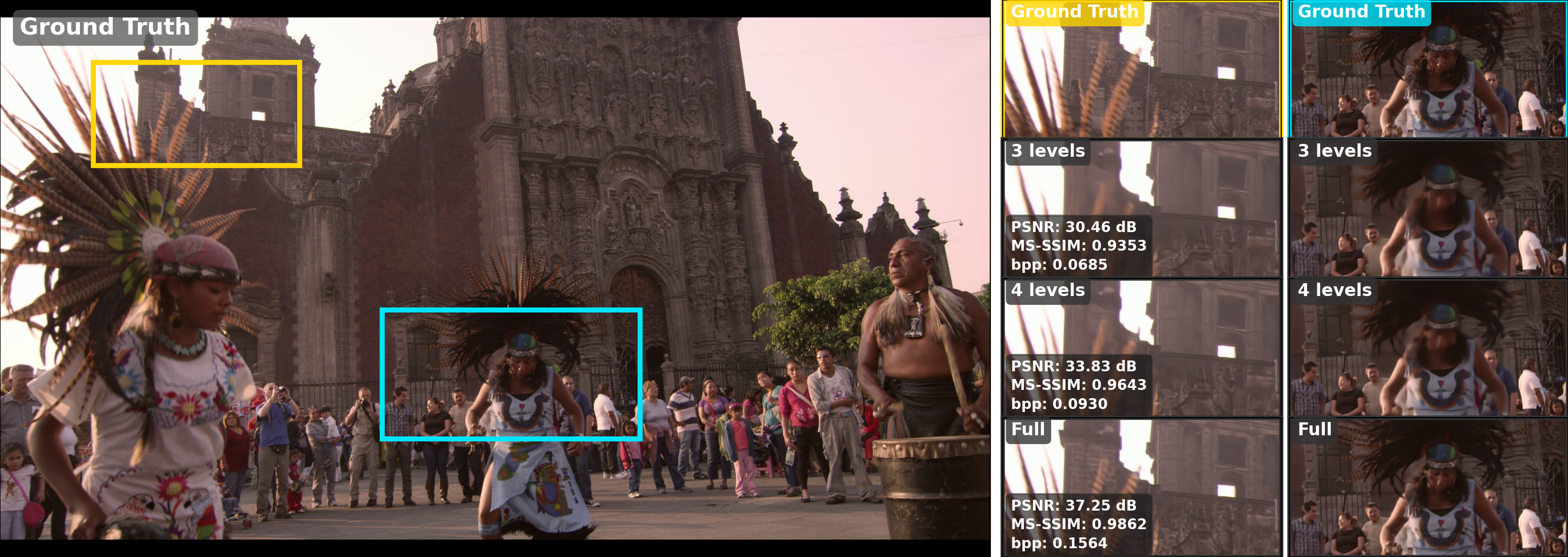}
        \caption{MCL-JCV - {videoSRC19}}
        \label{fig:scalable_mcl}
    \end{subfigure}
    \caption{Qualitative results of the inherent scalable coding capabilities of NVRC++. In these examples, our model achieves scalable coding by selectively reconstructing the video using 3, 4, or all 5 levels of the feature grids (progressing from the lowest to the highest resolution).}
    \label{fig:scalable_visuals}
\end{figure}

We provide visual examples of this scalable decoding in \cref{fig:scalable_visuals}, demonstrating the inherent versatility of our approach. Future work could explicitly target and further enhance this scalable coding performance.

\section{Extended visual comparisons}
\label{sec:qualitative_results}

\begin{figure}
    \centering
    \includegraphics[width=1\linewidth]{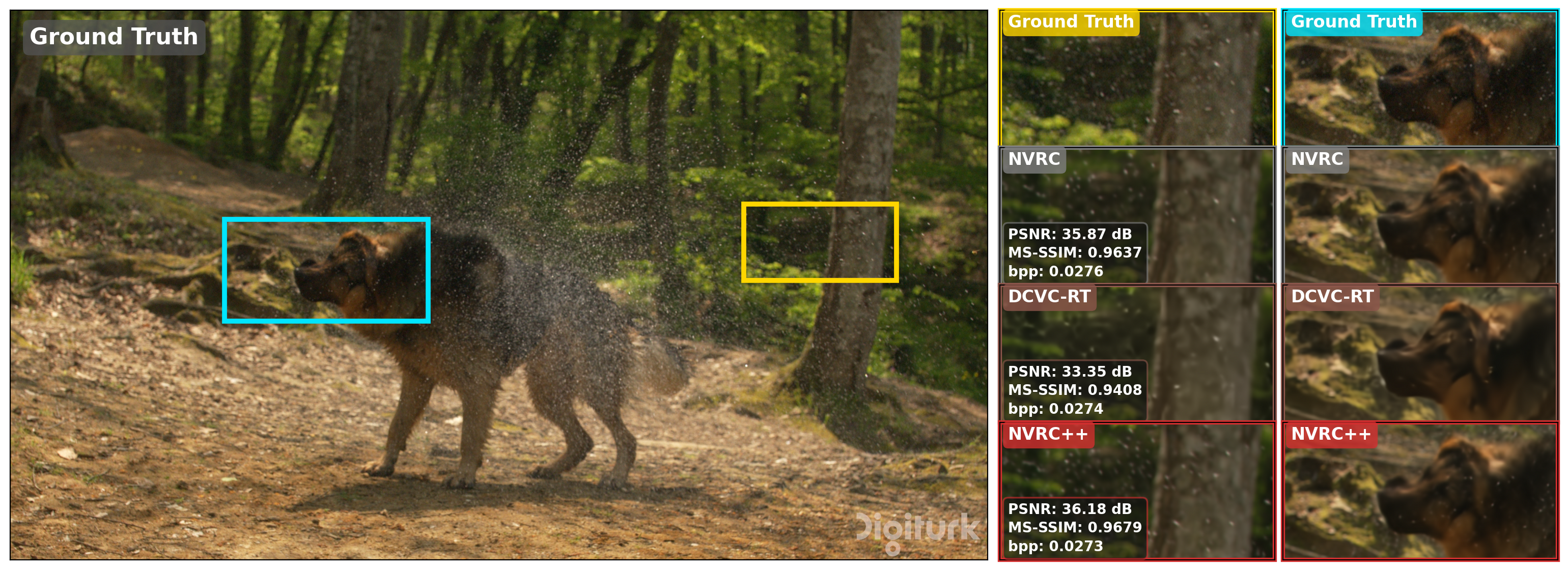}
    \caption{Extended visual comparisons on the UVG dataset (\textit{ShakeNDry}) \cite{mercat2020uvg}.}
    \label{fig:additional_visuals_1}
\end{figure}

\begin{figure}
    \centering
    \includegraphics[width=1\linewidth]{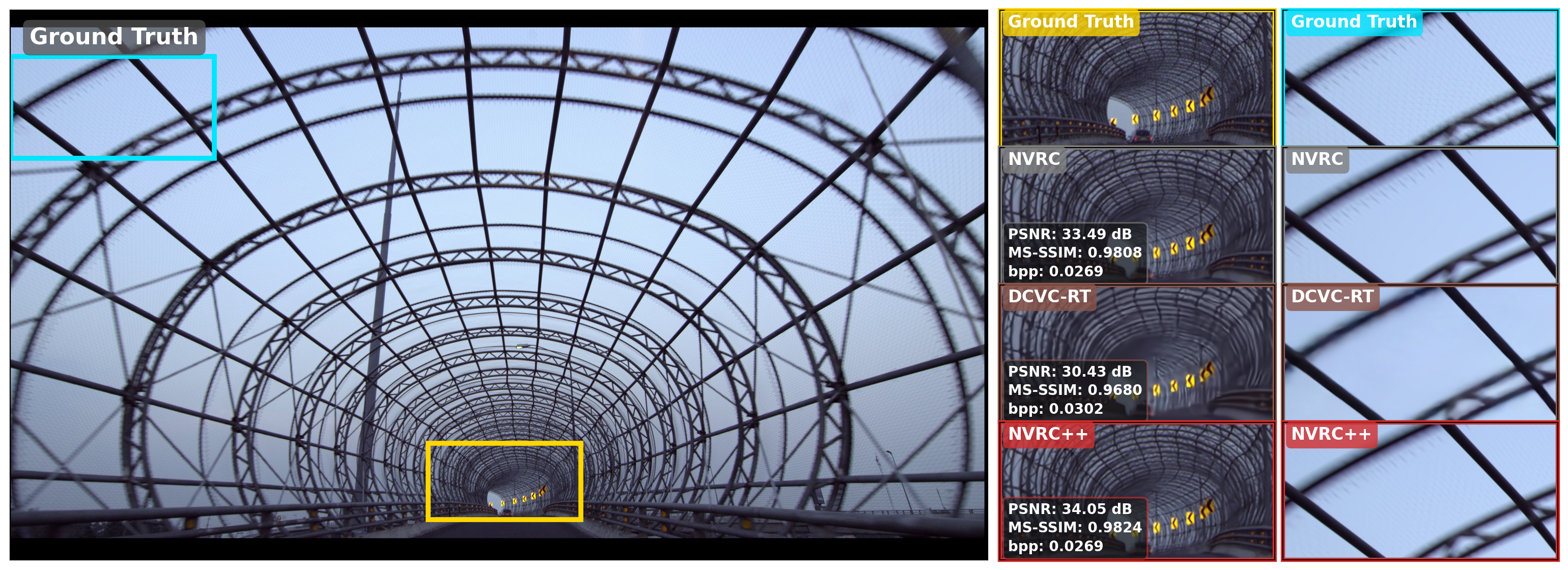}
    \caption{Extended visual comparisons: MCL-JCV dataset (\textit{videoSRC10}) \cite{wang2016mcl}.}
    \label{fig:additional_visuals_2}
\end{figure}

\begin{figure}
    \centering
    \includegraphics[width=1\linewidth]{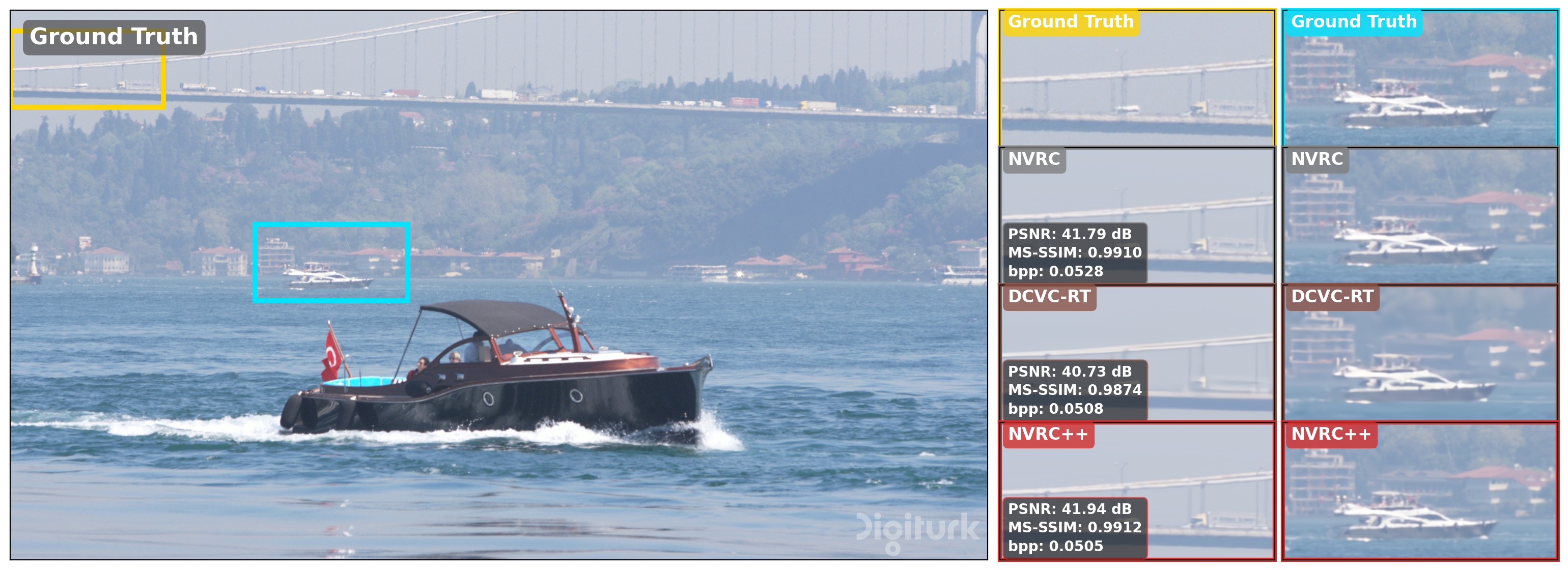}
    \caption{Extended visual comparisons: UVG dataset (\textit{Bosphorus}) \cite{mercat2020uvg}.}
    \label{fig:additional_visuals_3}
\end{figure}

\begin{figure}
    \centering
    \includegraphics[width=1\linewidth]{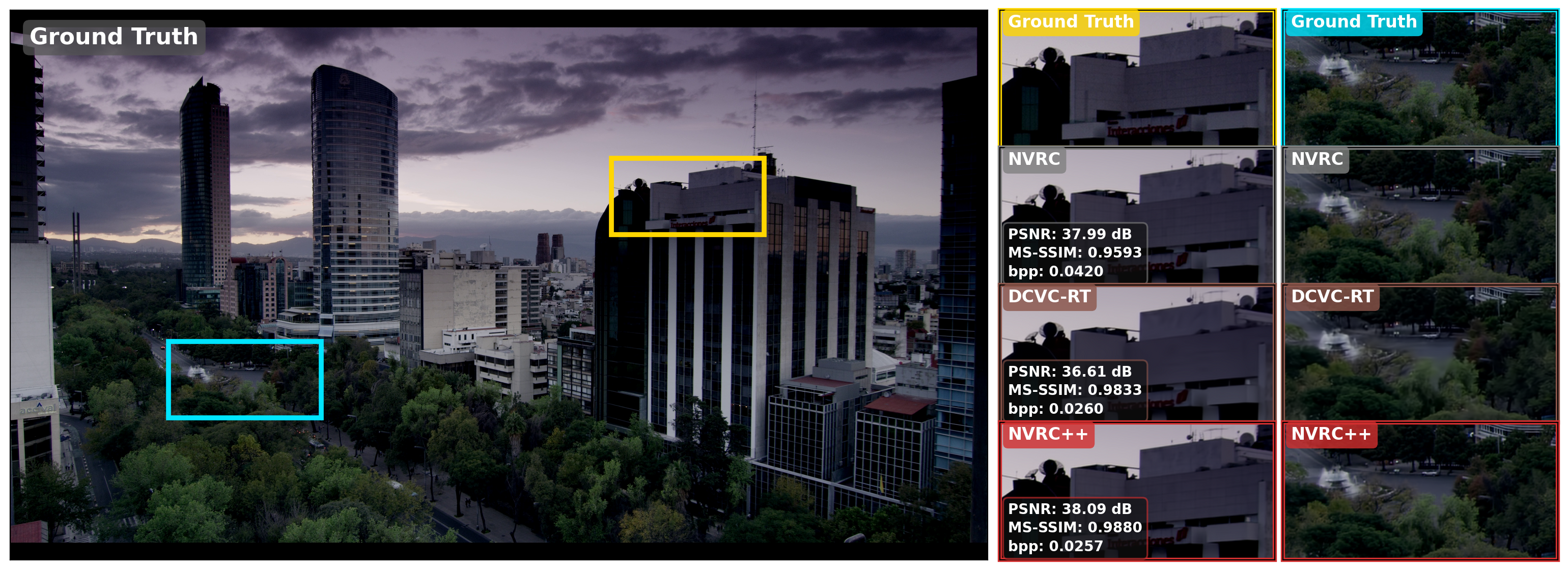}
    \caption{Extended visual comparisons: MCL-JCV dataset (\textit{videoSRC12}) \cite{wang2016mcl}.}
    \label{fig:additional_visuals_4}
\end{figure}

We provide additional visual results for evaluating the subjective quality in \cref{fig:additional_visuals_1,fig:additional_visuals_2,fig:additional_visuals_3,fig:additional_visuals_4}. Our results demonstrate that NVRC++ achieves superior visual quality compared to the baseline methods, including NVRC \cite{kwan2024nvrc} and DCVC-RT \cite{jia2025towards}.

\clearpage

\clearpage

\clearpage
\newpage

\end{document}